\documentclass[prl,twocolumn,citeautoscript,amsmath,superscriptaddress]{revtex4-1}

\usepackage{graphicx}
\usepackage{newtxtext}
\usepackage{epsfig}
\usepackage{amsmath,bbm,amssymb,bm,times}

\usepackage[colorlinks,linkcolor=blue,citecolor=blue]{hyperref}

\begin{document}

\title{Nonlinearity-induced topological phase transition characterized by the nonlinear Chern number}
\author{Kazuki Sone}
\email{sone@noneq.t.u-tokyo.ac.jp}
\affiliation{Department of Applied Physics, The University of Tokyo, 7-3-1 Hongo, Bunkyo-ku, Tokyo 113-8656, Japan}
\author{Motohiko Ezawa}
\affiliation{Department of Applied Physics, The University of Tokyo, 7-3-1 Hongo, Bunkyo-ku, Tokyo 113-8656, Japan}
\author{Yuto Ashida}
\affiliation{Department of Physics, University of Tokyo, 7-3-1 Hongo, Bunkyo-ku, Tokyo 113-0033, Japan}
\affiliation{Institute for Physics of Intelligence, University of Tokyo, 7-3-1 Hongo, Tokyo 113-0033, Japan}
\author{Nobuyuki Yoshioka}
\affiliation{Department of Applied Physics, The University of Tokyo, 7-3-1 Hongo, Bunkyo-ku, Tokyo 113-8656, Japan}
\affiliation{Theoretical Quantum Physics Laboratory, RIKEN Cluster for Pioneering Research (CPR), Wako-shi, Saitama
351-0198, Japan}
\affiliation{Japan Science and Technology Agency (JST), PRESTO, 4-1-8 Honcho, Kawaguchi, Saitama 332-0012, Japan}
\author{Takahiro Sagawa} 
\affiliation{Department of Applied Physics, The University of Tokyo, 7-3-1 Hongo, Bunkyo-ku, Tokyo 113-8656, Japan}
\affiliation{Quantum-Phase Electronics Center (QPEC), The University of Tokyo, 7-3-1 Hongo, Bunkyo-ku, Tokyo 113-8656, Japan}

\begin{abstract}
As first demonstrated by the characterization of the quantum Hall effect by the Chern number, topology provides a guiding principle to realize robust properties of condensed matter systems immune to the existence of disorder. The bulk-boundary correspondence guarantees the emergence of gapless boundary modes in a topological system whose bulk exhibits nonzero topological invariants. Although some recent studies have suggested a possible extension of the notion of topology to nonlinear systems such as photonics and electrical circuits, the nonlinear counterpart of topological invariant has not yet been understood. Here, we propose the nonlinear extension of the Chern number based on the nonlinear eigenvalue problems in two-dimensional systems and reveal the bulk-boundary correspondence beyond the weakly nonlinear regime. Specifically, we find the nonlinearity-induced topological phase transitions, where the existence of topological edge modes depends on the amplitude of oscillatory modes. We propose and analyze a minimal model of a nonlinear Chern insulator whose exact bulk solutions are analytically obtained and indicate the amplitude dependence of the nonlinear Chern number, for which we confirm the nonlinear counterpart of the bulk-boundary correspondence in the continuum limit. Thus, our result reveals the existence of genuinely nonlinear topological phases that are adiabatically disconnected from the linear regime, showing the promise for expanding the scope of topological classification of matter towards the nonlinear regime.
\end{abstract}
\maketitle
Topology is utilized to realize robust properties of materials that are immune to disorders \cite{Kane2005,Hasan2010}. A prototypical example of topological materials is the quantum Hall effect \cite{Klitzing1980, Thouless1982}, which was discovered in a two-dimensional semiconductor under a magnetic field. In such a two-dimensional system, the Chern number characterizes the topology of the band structure and the corresponding gapless boundary modes. This bulk-boundary correspondence lies at the heart of the robustness of topological devices utilizing boundary modes. Recent studies have also explored topological phenomena in a variety of platforms, such as photonics \cite{Lu2014}, electrical circuits \cite{Ningyuan2015}, ultracold atoms \cite{Jotzu2014}, fluids \cite{Yang2015}, and mechanical lattices \cite{Kane2013}.

While band topology has been well-explored in linear systems, nonlinear dynamics is ubiquitous in classical \cite{Boyd2003,Smirnova2020,Ota2020,Acebron2005,Strogatz2018,Marchetti2013} and interacting bosonic systems \cite{Gross1961,Pitaevskii1961}. For example, nonlinear interactions can naturally emerge in the mean-field analysis of bosonic many-body systems, as is seen in the Gross-Pitaevskii equations of ultracold atoms. Recent studies have also analyzed the nonlinear effects on topological edge modes \cite{Lumer2013,Bomantara2017,Harari2018,Zangeneh2019,Smirnova2020,Ota2020,Maczewsky2020,Lo2021,Jurgensen2021,Mochizuki2021,Fu2022,Mostaan2022} and revealed unique topological phenomena intertwined with solitons \cite{Chen2014,Leykam2016,Zhang2020,Ivanov2020,Mukherjee2021,Li2022a,Ezawa2022} and synchronization \cite{Kotwal2021,Sone2022,Wachtler2023}. Nonlinearity can further modify the conventional notion of topological phase; recent studies have revealed that one-dimensional systems can exhibit nonlinearity-induced topological phase transitions, where the existence of topological edge modes depends on the amplitude of the oscillatory modes \cite{Hadad2016,Darabi2019,Tuloup2020,Ezawa2021,Zhou2022}. While these previous studies have indicated the existence of topological edge modes in nonlinear systems, one cannot straightforwardly extend the topological invariants to nonlinear systems because they have no band structures. In addition, despite the advantage that nontrivial topology in two-dimensional systems requires no additional symmetries other than the $U(1)$ and translation symmetries as is the case for the quantum Hall effect, nonlinear topology in two-dimensional systems \cite{Lumer2013,Leykam2016,Zhang2020,Ezawa2022} is much less understood than that in one-dimensional systems.

\begin{figure*}
\includegraphics[width=140mm, bb=0 0 600 390,clip]{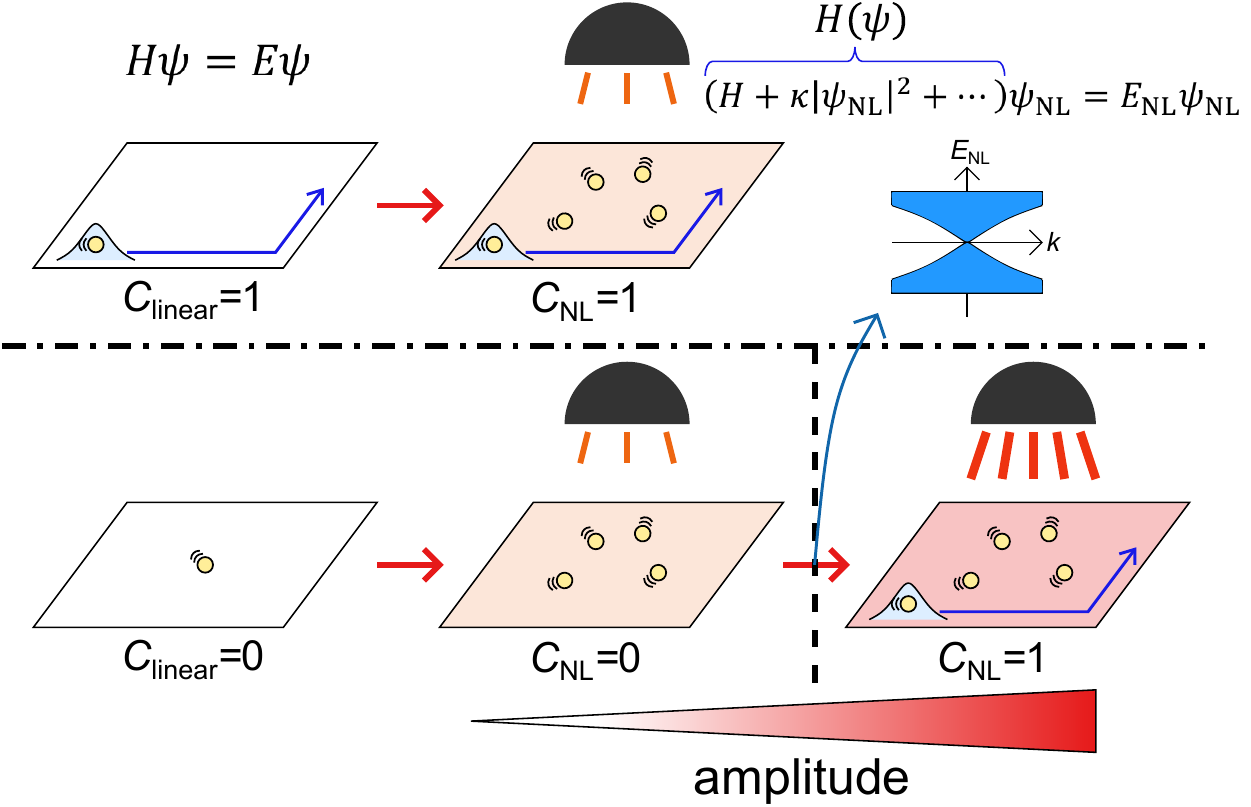}
\caption{\label{fig1}
{\bf Schematic of the nonlinear Chern insulators and the nonlinearity-induced topological phase transition.} While topology of a noninteracting linear system can be characterized by the Chern number that is computed from its eigenvectors, topology of a nonlinear system is classified by the nonlinear Chern number, which utilizes the nonlinear extension of the eigenequation. In weakly nonlinear regions (i.e., small amplitude), the nonlinear Chern number predicts the existence of edge modes corresponding to those in linear systems. Specifically, when nonlinear systems exhibit edge-localized steady states, both the nonlinear and linear Chern numbers are nonzero as shown in the upper part of the figure. If we inject higher energy into the system and consider the eigenmodes with large amplitudes, the nonlinear band structure can become gapless. At such a gapless point, nonlinearity-induced topological phase transition can occur, where topological boundary modes appear with the nonzero nonlinear Chern number. The nonlinearity-induced topological phases exhibit boundary modes that cannot be predicted from the linear Chern number. Therefore, such topological phases are genuinely unique to nonlinear systems. 
}
\end{figure*}

In this paper, we introduce the notion of the nonlinear Chern number and reveal its relation to the bulk-boundary correspondence. To define the nonlinear Chern number of two-dimensional systems, we consider the nonlinear extension of the eigenvalue problem \cite{Tuloup2020,Zhou2022,Li2022b} and make an analogy to band structures. While it is not obvious that the nonlinear eigenvalue problem elucidates the bulk-boundary correspondence in nonlinear topological insulators, we theoretically prove the bulk-boundary correspondence of the nonlinear Chern number in weakly nonlinear systems. Furthermore, in stronger nonlinear regimes where nonlinear terms are larger compared to the linear band gap, we find that the nonlinearity-induced topological phase transition can occur in two-dimensional systems as depicted in Fig.~\ref{fig1}. We construct a minimal model of a nonlinear Chern insulator, for which we can obtain exact bulk solutions and thus can analytically show the amplitude dependence of the associated nonlinear Chern number without any approximations. In the continuum limit of the model, we analytically show the nonlinear bulk-boundary correspondence, which states that the nonlinear Chern number predicts the existence of edge modes even under strong nonlinearity. In addition, we numerically confirm the bulk-boundary correspondence in the lattice model at the parameter where the lattice model well-approximates the behavior of the continuum model. We also discuss that the nonlinearity-induced edge modes can be observed in the quench dynamics in experimental setups such as nonlinear topological photonics. Since the nonlinearity-induced topological phases are disconnected from the linear limits under adiabatic deformations, our results show the existence of genuinely nonlinear topological phases.

The scope of this paper applies to a broad class of two-dimensional systems with $U(1)$-gauge and spatial translation symmetries, which can be realized in a variety of experimental setups. As is the case that the Thouless-Kohmoto-Nightingale-den Nijs formula \cite{Thouless1982} has triggered the research of a variety of topological materials, the nonlinear Chern number is expected to open up the research stream of nonlinear topological materials including their systematic classification. From the experimental point of view, one can realize nonlinear Chern insulators with the $U(1)$-gauge symmetry in, e.g., photonics \cite{Lumer2013,Harari2018,Smirnova2020,Ota2020,Maczewsky2020,Jurgensen2021,Mostaan2022,Leykam2016,Zhang2020,Ivanov2020,Mukherjee2021,Li2022a}, ultracold atoms \cite{Gross1961,Pitaevskii1961,Mostaan2022}, and electrical circuits \cite{Ningyuan2015,Kotwal2021,Sone2022}, where both linear band topology and nonlinear effects have been investigated. In particular, since Kerr nonlinearity \cite{Boyd2003} is fairly common in photonic systems, it should be possible to extend the current topological photonic devices to nonlinear ones.

{\it Nonlinear eigenvalue problem and nonlinear Chern number.---}
The conventional band topology is based on eigenvalue problems in linear systems. To define a topological invariant, one should consider eigenvalues $E_j(\mathbf{k})$ and eigenvectors $|\psi_j(\mathbf{k})\rangle$ of the Bloch Hamiltonian $H(\mathbf{k})$ corresponding to the Fourier component of the real-space Hamiltonian of a periodic system. Then, one can define the Chern number as $C_j = (1/2\pi i) \int \nabla_{\mathbf{k}} \times \langle \psi_j(\mathbf{k})| \nabla_{\mathbf{k}} |\psi_j(\mathbf{k}) \rangle d^2\mathbf{k}$. The Chern number is unchanged under the continuous deformation of the Hamiltonian unless the Bloch Hamiltonian is degenerate. In this sense, the band topology is robust against perturbations, which is of practical use to realize topological edge modes in experimental setups. Thus, to define the topological invariant in nonlinear systems, it is essential to extend the notion of band structures.

\begin{figure*}
  \includegraphics[width=140mm,bb=0 0 530 430,clip]{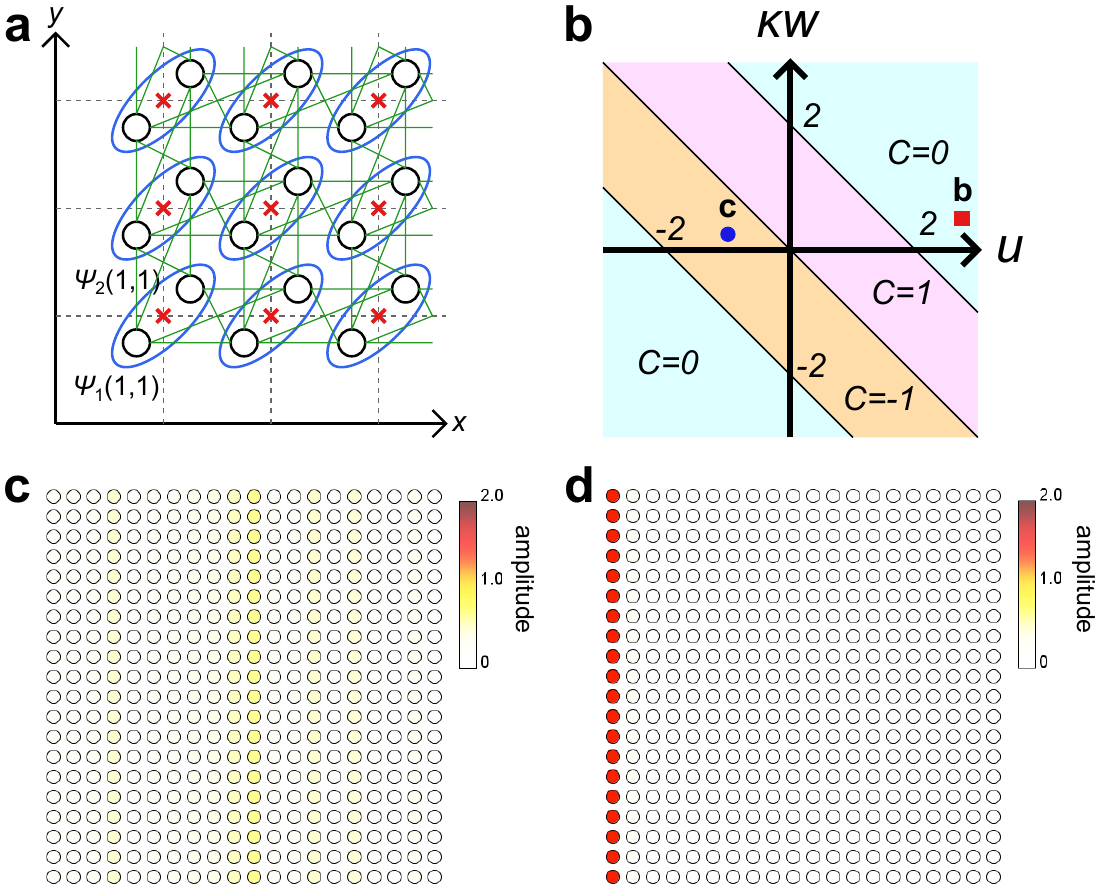}
  \caption{\label{fig2}
{\bf Phase diagram and dynamics of the nonlinear Qi-Wu-Zhang (QWZ) model.} {\bf a,} The figure shows a schematic of the nonlinear QWZ model. The model has two sublattices (the black circles) at each lattice point encircled by the blue ellipse. The green lines represent the linear couplings. We use the notation $\Psi_i(x,y)$ to represent the state variable at each sublattice, where $(x,y)$ is the location of the representative point of each lattice point denoted by the red cross. {\bf b,} We analytically obtain the phase diagram of the prototypical model of nonlinear topological insulators. The horizontal axis represents the parameter of the mass term and the vertical axis corresponds to the strength of the nonlinearity. The color of each separated region represents the difference in the nonlinear Chern number. {\bf c,} We numerically show the absence of the edge modes in the topologically-trivial parameter region. We simulate the dynamics of the prototypical model of a nonlinear Chern insulator starting from an initial state localized at the left edge. We impose the open boundary condition in the $x$ direction and the periodic boundary condition in the $y$ direction. The figure shows the snapshot at $t=1$. The color shows the absolute value of the components of the state vector at each site. The parameters used are $u=3$, $\kappa=0.1$, and $w=1$, which corresponds to the red square in panel {\bf b}. {\bf d,} We numerically show the existence of the long-lived localized state in the weakly nonlinear topological insulator. The figure shows the snapshot of the simulation at $t=1$. The sites at the left edge show large amplitudes, which indicates the existence of the edge-localized state. The parameters used are $u=-1$, $\kappa=0.1$, and $w=1$, which corresponds to the blue circle in panel {\bf b}.}
\end{figure*}

We here consider the nonlinear extension of the eigenvalue equations \cite{Tuloup2020,Zhou2022,Li2022b} and define the nonlinear Chern number by using the nonlinear eigenvectors. We start from the general nonlinear dynamics,
\begin{equation}
i\partial_t \Psi_j (\mathbf{r}) = f_j(\Psi;\mathbf{r}),\label{nonlinear_Eq}
\end{equation}
where $\Psi_j (\mathbf{r})$ is the state variable and $f_j(\cdot;\mathbf{r})$ is a nonlinear function of the state vector $\Psi$. In lattice systems, we use the notation that $\mathbf{r}$ is a representative point in each unit cell of the lattice as shown in Fig.~\ref{fig2}a. Then, $j$ represents the internal degrees of freedom that include, e.g., sublattices and effective spin degrees of freedom. When we consider continuum systems, $\mathbf{r}$ should simply represent the location and $j$ corresponds to the internal degrees of freedom such as spins. For example, the Gross-Pitaevskii equation in the continuous space is given by $f(\Psi;\mathbf{r})=-\nabla^2\Psi(\mathbf{r})/(2m)+V\Psi(\mathbf{r})+(4\pi a/m)|\Psi(\mathbf{r})|^2\Psi(\mathbf{r})$ with $V$ being a potential and $m$ and $a$ being constants, and the nonlinear function $f(\cdot;\mathbf{r})$ depends on the $\Psi(\mathbf{r})$ and its derivative and has no internal degrees of freedom. Since the quantum Hall system has the $U(1)$ and translation symmetries, we impose them on the nonlinear equation to study the analogy of such a prototypical topological insulator. Concretely, the $U(1)$ symmetry is represented as $f_j(e^{i\theta}\Psi;\mathbf{r})=e^{i\theta}f_j(\Psi;\mathbf{r})$, which is satisfied in, e.g., the Kerr-like nonlinearity $\kappa |\Psi_j(\mathbf{r})|^2\Psi_j(\mathbf{r})$ \cite{Boyd2003}. The translational symmetry in lattice systems is defined as $f_j(\Psi;\mathbf{r}+\mathbf{a}) = f_j(\Phi;\mathbf{r})$ with $\mathbf{a}$ being a lattice vector and $\Phi$ being translated state variables $\Phi_j(\mathbf{r}) = \Psi_j(\mathbf{r}+\mathbf{a})$. The translational symmetry in continuum systems is also defined in the same equations, while $f_j(\Phi;\mathbf{r})$ still remains $\mathbf{r}$ dependence due to, e.g., the periodic potential. We also focus on conservative dynamics analogous to Hermitian Hamiltonians where the sum of squared amplitudes $\sum_{j,\mathbf{r}}|\Psi_j(\mathbf{r})|^2$ is preserved. 

Corresponding to the nonlinear dynamical system in Eq.~\eqref{nonlinear_Eq}, the nonlinear eigenvector and eigenvalue are defined as the state vector $\Psi$ with components $\Psi_j(\mathbf{r})$ and the constant $E$ that satisfy 
\begin{equation}
f_j(\Psi;\mathbf{r}) = E\Psi_j (\mathbf{r}). \label{NLeigen}
\end{equation}
We term the equation as a nonlinear eigenequation and analyze its bulk-boundary correspondence below. We note that we can regard the nonlinear eigenvector as a periodically oscillating steady state $\Psi_j(\mathbf{r};t)=e^{-iEt}\Psi_j(\mathbf{r})$ of the nonlinear system when the eigenvalue is real. Since the sum of the squared amplitude is conserved under the $U(1)$ symmetry, we here focus on special solutions with the fixed sums of squared amplitudes $\sum_{j,\mathbf{r}} |\Psi_j(\mathbf{r})|^2$ (resp. $\sum_{j} \int d^n\mathbf{r} |\Psi_j(\mathbf{r})|^2$) in lattice (resp. continuum) systems, where we take the sum both on the location and the internal degrees of freedom. 

To extend the Chern number to nonlinear systems, we introduce the eigenvalue problems in the wavenumber space, which is analogous to the linear eigenequation of the Bloch Hamiltonian. In a lattice system with the translation symmetries, we assume an ansatz state \cite{Tuloup2020} that we name the Bloch ansatz: $\Psi_j(\mathbf{r})=e^{i\mathbf{k}\cdot\mathbf{r}}\psi_j(\mathbf{k})$. We note that in linear systems the Bloch theorem guarantees that every eigenvector is given by the form of the Bloch ansatz.  On the other hand, in nonlinear systems, there can be nonlinear eigenvectors out of the description of the Bloch ansatz, including bulk-localized ones. Despite the existence of such localized modes, we here only focus on nonlinear bulk eigenvectors described by the Bloch ansatz and show that even such periodic bulk solutions can exhibit unique topological phenomena to nonlinear systems, i.e., the nonlinearity-induced topological phase transition. Under this ansatz and the $U(1)$ symmetry, one can rewrite the nonlinear eigenequation as $f_j(\mathbf{k},\psi(\mathbf{k})) = E(\mathbf{k})\psi_j(\mathbf{k})$ parametrized by $\mathbf{k}$ (see Supplementary Information for the detailed derivation). 

To capture the nonlinearity-induced topological phase transition depending on amplitudes, we focus on a special solution of the nonlinear eigenvector at each $\mathbf{k}$ whose sum of the squared amplitudes $\sum_j |\psi_j(\mathbf{k})|^2=w$ is fixed independently of the wavenumber $\mathbf{k}$. We note that the assumption of such fixed-amplitude Bloch-ansatz solutions is consistent with the perturbation calculation of the nonlinear eigenvectors (see Supplementary Information for the detail). By using fixed-amplitude nonlinear eigenvectors, we define the nonlinear Chern number as 
\begin{equation}
C_{{\rm NL}}(w) = \frac{1}{2\pi i} \int \nabla_{\mathbf{k}} \times \left(\frac{1}{\sqrt{w}}\langle \psi(\mathbf{k})|\right) \nabla_{\mathbf{k}} \left(\frac{1}{\sqrt{w}}|\psi(\mathbf{k}) \rangle\right) d^2\mathbf{k}.
\end{equation}
We note that this definition reduces to the conventional linear Chern number if $f$ defined in Eq.~\eqref{NLeigen} is a linear function. It is also noteworthy that since special solutions of nonlinear eigenvectors should exist at arbitrary $w$ in ordinary nonlinear systems, we can define the nonlinear Chern number at any positive $w$ except for gap-closing points. One can prove that the nonlinear Chern number is an integer by embedding the nonlinear eigenvectors into an eigenspace of a linear Bloch Hamiltonian (see also Supplementary Information). Since the eigenvector can be changed by the amplitude $w$, the nonlinear Chern number also depends on $w$ as shown in Fig.~\ref{fig1}. Therefore, the nonlinear Chern number can predict the nonlinearity-induced topological phase transition by the change of the amplitude in nonlinear systems, which is a qualitatively novel phenomenon absent in linear systems \cite{Maczewsky2020,Hadad2016,Darabi2019,Tuloup2020,Mochizuki2021,Ezawa2021,Zhou2022}.

In continuum systems with a periodic potential, the Bloch ansatz should read $\Psi_j(\mathbf{r})=e^{i\mathbf{k}\cdot\mathbf{r}}\psi_j(\mathbf{k},\mathbf{r})$ with $\psi_j(\mathbf{k},\mathbf{r})$ being a periodic function of $\mathbf{r}$ whose period is equal to that of the periodic potential. Then, the wavenumber-space representation of the nonlinear eigenequation becomes $f_j(\mathbf{k},\psi(\mathbf{k});\mathbf{r}) = E(\mathbf{k})\psi_j(\mathbf{k},\mathbf{r})$, and the nonlinear Chern number can be written as 
$C_{{\rm NL}} = 1/(2\pi i) \sum_j \int_S d^2\mathbf{r} \int \nabla_{\mathbf{k}} \times \left[ (\psi_j(\mathbf{k},\mathbf{r})/\sqrt{w}) \nabla_{\mathbf{k}} (\psi_j(\mathbf{k},\mathbf{r})/\sqrt{w} ) \right] d^2\mathbf{k}$, where $S$ represents the unit cell of the periodic system. The squared amplitude is defined as $w = \sum_j \int_S d^2\mathbf{r} |\psi_j(\mathbf{k},\mathbf{r})|^2$ in this continuum case.

Compared to a previous study of the nonlinear topological invariant in one-dimensional systems \cite{Zhou2022}, the nonlinear Chern number has no higher-order correction terms. This is because we assume the $U(1)$ symmetry which guarantees that one can describe periodic steady states by using a single frequency mode. Therefore, there is no need to consider the multi-mode expansion of nonlinear eigenvectors that leads to the nonlinear collection terms. In fact, the one-dimensional topological invariant defined in the previous study is also reduced to the conventional form under the $U(1)$ symmetry, which is consistent with our results.

We note that the Bloch ansatz does not describe bulk-localized solutions that can be obtained in strongly nonlinear systems. Since such strong nonlinearity can also generate edge-localized modes \cite{Ezawa2022}, it is not straightforward to identify whether the edge modes originate from the bulk topology or the nontopological nonlinear effects, which makes the bulk-boundary correspondence unclear in strongly nonlinear regimes. Therefore, in the following sections, we mainly focus on weakly and more strongly nonlinear systems where the nonlinear terms are smaller than the linear terms (see also Supplementary Information for the Bloch-wave-like solutions of the bulk modes in this parameter region).

\begin{figure*}
  \includegraphics[width=140mm,bb=0 0 1245 305,clip]{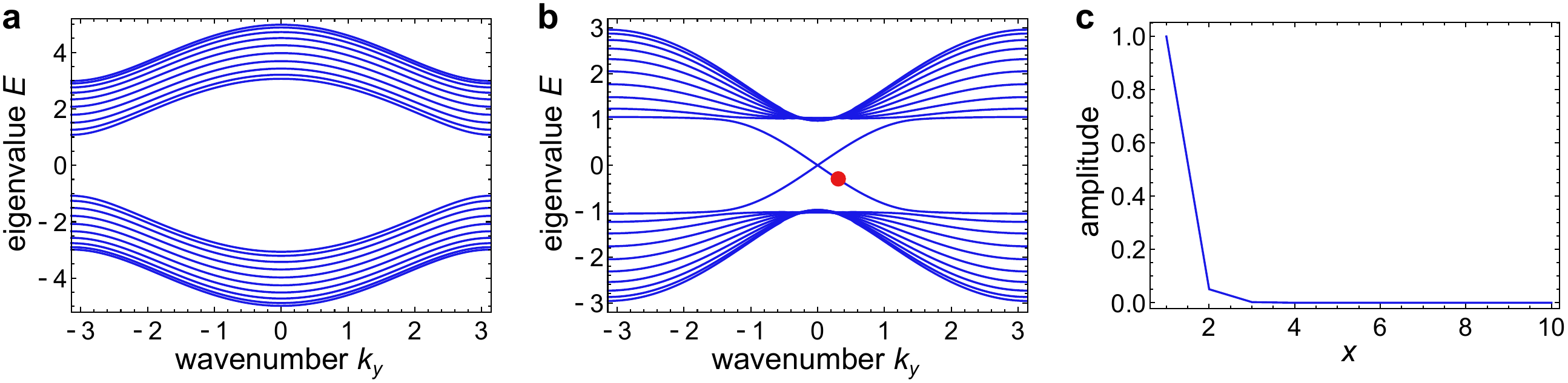}
  \caption{\label{fig3} 
  {\bf Nonlinear band structure and edge modes of the weakly nonlinear QWZ model.} {\bf a,} We numerically calculate the nonlinear band structure of the topologically trivial system under the open boundary condition in the $x$ direction and the periodic boundary condition in the $y$ direction. One can confirm the absence of gapless modes. The parameters used are  $u=3$, $\kappa=0.1$, and $w=1$. {\bf b,} Nonlinear band structure of the topologically nontrivial system is numerically calculated. There are gapless modes that connect the upper and lower bulk bands. The parameters used are $u=-1$, $\kappa=0.1$, and $w=1$. {\bf c,} The spatial distribution of the gapless mode is presented. The eigenvalue of the localized mode corresponds to the red circle in the band structure in panel {\bf b}.
  }
\end{figure*}

{\it Nonlinear Chern number calculated from exact solutions of a lattice model.---}
To investigate the bulk-boundary correspondence, i.e., the correspondence between the nonzero nonlinear Chern number and the existence of the edge-localized steady state, we propose and analyze the nonlinear extension of the Qi-Wu-Zhang (QWZ) model \cite{Qi2006} defined in the real space (see Supplementary Information for the detail of the real-space description of the model). By using the Bloch ansatz, we rigorously obtain its wavenumber-space description,
\begin{widetext}
\begin{eqnarray}
f(\mathbf{k},\psi(\mathbf{k})) = \left(
  \begin{array}{cc}
   u + \kappa w + \cos k_x+\cos k_y  & \sin k_x + i\sin k_y \\
   \sin k_x - i\sin k_y & -(u + \kappa w +\cos k_x+\cos k_y)
  \end{array}
  \right) \left(
  \begin{array}{c}
   \psi_1(\mathbf{k}) \\
   \psi_2(\mathbf{k})
  \end{array}
  \right), \label{model-wavenumber}
\end{eqnarray}
\end{widetext}
where $w$ is the squared amplitude $w=|\psi_1(\mathbf{k}) |^2+|\psi_2(\mathbf{k}) |^2$. $u$ and $\kappa$ are dimensionless parameters of linear and nonlinear mass terms. We here introduce the staggered Kerr-like nonlinearity $\pm \kappa w$ to the linear Chern-insulator model \cite{Qi2006}.

To calculate the nonlinear Chern number, we focus on special solutions where the squared amplitude $w$ is fixed independently of the wavenumber $\mathbf{k}$. Then, we can regard Eq.~\eqref{model-wavenumber} as a linear equation and thus can diagonalize it as the linear QWZ model. Analytically diagonalizing the right-hand side of Eq.~\eqref{model-wavenumber}, we obtain the exact bulk solutions of the nonlinear eigenvalues and eigenvectors. Then, using the exactly obtained nonlinear eigenvectors, we calculate the nonlinear Chern number and obtain the phase diagram in Fig.~\ref{fig2}b (see also Supplementary Information for the detail of the calculation and Supplementary Information for the numerical confirmation). The amplitude dependence of the nonlinear Chern number indicates the existence of the nonlinearity-induced topological phase transition in the nonlinear QWZ model. We note that the above nonlinear eigenvectors are exact solutions of the nonlinear QWZ model under the periodic boundary conditions, and thus our result reveals the existence of the nonlinearity-induced topological phase transition without any approximations. Such an analytical demonstration of the nonlinearity-induced topological phase transition is achieved by considering nonlinear equations with the form,
\begin{equation}
i\partial _{t} \psi _{n}\left( \mathbf{k};t\right)
=\sum_{m}f_{nm}\left( \sum_j \left\vert \psi _{j}\left( \mathbf{k};t\right) \right\vert ^{2}\right) \psi _{m}\left( \mathbf{k};t\right),\label{general_nonlinear_eq}
\end{equation}
with $\psi _{j}$ and $\mathbf{k}$ being the state variables and the wavenumber (see also Supplementary Information), while the addition of off-diagonal or non-uniform diagonal nonlinear terms prevents us to obtain the exact solutions. It is also noteworthy that the obtained exact solutions are consistent with the results of the perturbation or self-consistent calculations of the nonlinear eigenvalue problem (see also Supplementary Information).

{\it Bulk-boundary correspondence in weakly nonlinear systems.---} 
We first numerically confirm the bulk-boundary correspondence in weakly nonlinear systems. We simulate the dynamics of the nonlinear QWZ model (Eq.~\eqref{model-wavenumber}) with weak nonlinearity where the nonlinear Chern number is the same as that in the linear limit $\kappa w \rightarrow 0$. In the topological phase $C_{\rm NL}=\pm1$ (Fig.~\ref{fig2}d), we find a long-lived localized state that corresponds to a topological edge mode in the QWZ model. Meanwhile, in the case of $C_{\rm NL}=0$ (Fig.~\ref{fig2}c), the edge-localized initial state is spread to the bulk, which indicates the absence of edge modes. We also confirm the bulk-boundary correspondence from the perspective of the nonlinear band structure as shown in Fig.~\ref{fig3} (see Supplementary Information for the details of the numerical method). 

In fact, the bulk-boundary correspondence between the nonlinear Chern number and the gapless edge modes can be established in general weakly nonlinear systems. We mathematically show the bulk-boundary correspondence under weak nonlinearity compared to the linear band gap. We describe the detail of the theorem and its proof in Supplementary Information.

\begin{figure*}
  \includegraphics[width=140mm,bb=0 0 1570 715,clip]{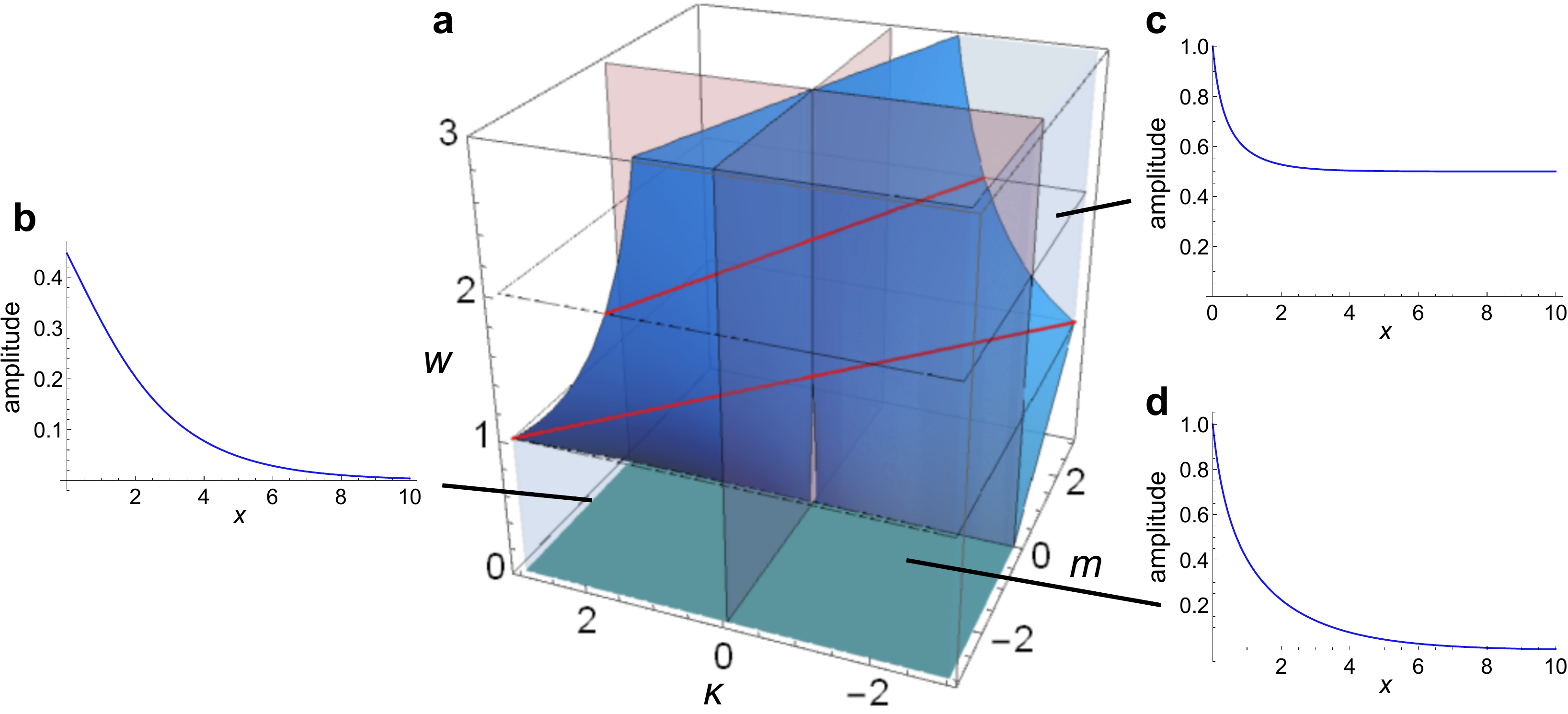}
  \caption{\label{fig4} 
  {\bf Phase diagram and localized modes of the nonlinear Dirac Hamiltonians.} {\bf a,} We illustrate the phase diagram of the nonlinear Dirac Hamiltonian, which demonstrates the nonlinear bulk-boundary correspondence in the continuum model. The vertical axis represents the amplitude, and the horizontal axes correspond to the parameters of the nonlinear Dirac Hamiltonian. The blue curved surface shows the phase boundary that separates a trivial phase without boundary modes and a topological phase exhibiting localized modes at the left boundary. The red surfaces present the boundaries where the sign of the parameters of the Dirac Hamiltonian changes. The red lines show the phase boundaries in the surfaces of $w=1$ and $w=2$. In the linear limit ($w=0$), the topological phases are separated by the $m=0$ axis, while the nonlinearity modifies the boundary of the topological phases. {\bf b-d,} Each panel shows the representative shape of the localized mode in each of topologically nontrivial parameter regions. {\bf b,} When $m$ is negative and $\kappa$ is positive, we obtain a localized mode in the small-amplitude region, which is regarded as a counterpart of a conventional topological edge mode. We set $m$, $\kappa$, and $D$ as $m=-0.5$, $\kappa=1$, and $D=3$. {\bf c,} When $m$ is positive and $\kappa$ is negative, the nonlinear Dirac Hamiltonian exhibits the nonlinearity-induced topological phase transition. We obtain an unconventional localized mode if the amplitude is larger than a critical value. In this localized mode, there exist nonvanishing amplitudes even in the limit of $x\rightarrow\infty$. Therefore, the localized mode can be unphysical in a truly semi-infinite system because we cannot normalize such a nonvanishing mode, while it still exists and is physically relevant in experimentally realizable finite systems. We set $m$, $\kappa$, and $D$ as $m=0.5$, $\kappa=-2$, and $D=-3$. {\bf d,} When both $m$ and $\kappa$ are negative, a localized mode appears independently of the amplitude as in linear topological insulators. We set $m$, $\kappa$, and $D$ as $m=-0.5$, $\kappa=-1$, and $D=3$.}
\end{figure*}

{\it Nonlinearity-induced topological phase transition by stronger nonlinearity.---}
We next show that nonlinearity-induced topological phase transitions occur in the stronger nonlinear regime, where the nonlinear Chern number becomes nonzero and topological edge modes appear at a critical amplitude. To analyze the behavior of topological edge modes, we derive the effective theory of the low-energy dispersion of the nonlinear QWZ model around the critical amplitude. For example, if we focus on the critical amplitude $w_c=(-2-u)/\kappa$, the nonlinear band structure of the model closes the gap at $(k_x,k_y)=(0,0)$. Then, around the critical amplitude and the gap closing point, $w\sim w_c$ and $(k_x,k_y)\sim(0,0)$, we remain the leading order term of the wavenumber-space description of the nonlinear QWZ model. Finally, by substituting the wavenumbers with the derivative, we derive the real-space description of the continuum model
\begin{widetext}
\begin{eqnarray}
i\partial_t \Psi(\mathbf{r}) &=& \hat{H}(\Psi(\mathbf{r})) \Psi(\mathbf{r}), \\
\hat{H}(\Psi(\mathbf{r})) &=& \left(
  \begin{array}{cc}
   m + \kappa (|\Psi_1(\mathbf{r})|^2 + |\Psi_2(\mathbf{r})|^2) & -i\partial_x + \partial_y \\
   -i\partial_x - \partial_y & -m - \kappa (|\Psi_1(\mathbf{r})|^2 + |\Psi_2(\mathbf{r})|^2)
  \end{array}
  \right), \label{nonlinearDirac}
\end{eqnarray}
\end{widetext}
with $m$ being $m=u+2$ and $\psi(x,y)=(\psi_1(x,y),\psi_2(x,y))^T$ being the state-vector function at the location $(x,y)$. This state-dependent Hamiltonian has a similar structure to the Dirac Hamiltonian except for the nonlinear mass term $\kappa (|\Psi_1|^2 + |\Psi_2|^2)$, and thus we term it the nonlinear Dirac Hamiltonian. Starting from other critical amplitudes $w_c=-u/\kappa$ and $w_c=(2-u)/\kappa$, one can derive similar state-dependent Hamiltonians (see also Supplementary Information). In general, the nonlinear Dirac Hamiltonian should describe the low-energy dispersion of nonlinear topological insulators, and thus its localized modes unveil the existence of topological edge modes in various continuum systems. 

By using the Bloch ansatz $\Psi_i(x,y)=\psi_i(\mathbf{k}) \exp(i(k_x x+k_y y))$ (without explicit $(x,y)$ dependence because of the continuous translational symmetry), one can determine the nonlinear Chern number of the nonlinear Dirac Hamiltonian as 
\begin{eqnarray}
C_{\rm NL} = \begin{cases}
\frac{1}{2} & (m+\kappa w > 0) \\
-\frac{1}{2} & (m+\kappa w< 0)
\end{cases}, \label{Chern-Dirac}
\end{eqnarray}
where $w=\int_S dS [|\Psi_1(x,y)|^2 + |\Psi_2(x,y)|^2]/|S|$ is the average squared amplitude of plain waves in this nonlinear system. We note that the nonlinear Chern number of the nonlinear QWZ model corresponds to the sum of those of the nonlinear Dirac Hamiltonians obtained from the expansion around the gap-closing points $(k_x,k_y)=(0,0),\ (0,\pi),\ (\pi,0),\ (\pi,\pi)$. 

\begin{figure}
\includegraphics[width=70mm, bb=0 0 340 210,clip]{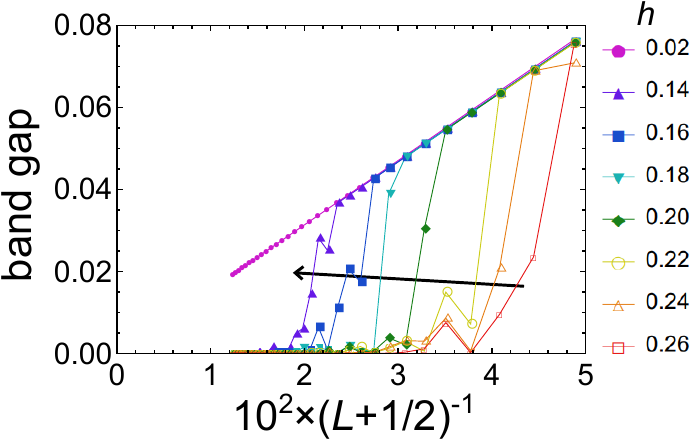}
\caption{\label{fig5} {\bf The bulk-boundary correspondence for the lattice model in the continuum limit.} We calculate the band gaps of the nonlinear QWZ model at different sizes and lattice constants. We impose the open boundary condition in the $x$ direction and the periodic boundary condition in the $y$ direction. We fix the parameters $m=-1$, and $\kappa w=1.0$, where the nonlinear band gap is closed in the corresponding continuum model. The band gaps versus $10^2/(L+1/2)$ ($L$ is the size of the system) are plotted. The legend shows the lattice constants $h$. We can confirm that the size of the gap decreases as the system size becomes larger. However, if the lattice constant is so large that the discretized system cannot imitate the behavior of the continuum nonlinear Dirac Hamiltonian, we find a sudden decrease in the size of the band gap. This sudden decrease contradicts the estimation that the band gap is proportional to $1/(L+1/2)$ and indicates the existence of irregular gapless modes. Since such gapless modes appear at smaller amplitudes than the critical point of the nonlinear Chern number, the numerical result indicates that the bulk-boundary correspondence can be ensured only after taking both the continuum and thermodynamic limits.}
\end{figure}

We analytically show that the Chern number predicts the existence of localized modes in continuum systems by calculating the localized modes of the nonlinear Dirac Hamiltonian. We assume the ansatz $\Psi_i(x,y) = e^{ik_y y}\Psi'_i(x)$ that is periodic in the $y$ direction and consider the $x>0$ region with the open boundary at $x=0$. We calculate the localized mode with the wavenumber $k_y$ and the eigenvector $E=k_y$. Constructing an analogy to the linear case, one can use an ansatz $(\Psi_1(x,y),\Psi_2(x,y))^T=e^{ik_y y}\phi(x)(1/\sqrt{2},-i/\sqrt{2})^T$ to describe the localized mode. Then, $\phi(x)$ should satisfy $\partial_x \phi(x) = m \phi(x) + \kappa |\phi(x)|^2 \phi(x)$. We can analytically obtain the solution of this equation,
\begin{equation}
 \phi(x) = e^{i\theta} \sqrt{\frac{1}{-(\kappa / m) + De^{-2mx}}}, \label{localized-mode}
\end{equation}
where $D$ is the integral constant and $-(\kappa / m) + De^{-2mx}$ must be positive. Combining $w$ and $D$ by the averaged squared amplitude $w=\int_0^L dx \left[1/ \{-(\kappa / m) + De^{-2mx}\} \right] / L = \ln [(1-\kappa/Dm)/(1-e^{2Lm}\kappa/Dm)]/(2\kappa L)$ for a fixed $L$ (inversely, $D=[\kappa(1-e^{2L(m+\kappa w)})]/[m(1-e^{2\kappa wL})]$), we find the bulk-boundary correspondence, i.e., correspondence between the positive nonlinear Chern number in Eq.~\eqref{Chern-Dirac} and the existence of a left-localized mode in Eq.~\eqref{localized-mode}. We note that $L$ can take an arbitrary value because the local amplitude $w(x) = 1/ \sqrt{-(\kappa / m) + De^{-2mx}}$ of the left-localized mode also exists in the topological parameter region, $m+\kappa w(x) < 0$.

Figure \ref{fig4} summarizes the behaviors of gapless modes in the nonlinear Dirac Hamiltonian at different parameters. In the case of $m<0$, where the Chern number is $C=1/2$ in the linear limit, we obtain the localized states at the left side as in the linear case. These localized states are consistent with the bulk-boundary correspondence in weakly nonlinear systems shown in the previous section. If we consider $m>0$, $\kappa<0$, we still obtain the localized state, while the amplitude remains to be $\sqrt{|m/\kappa|}$ at the limit of $x\rightarrow \infty$. The residual amplitude $\sqrt{|m/\kappa|}$ corresponds to the transition point of the nonlinear Chern number that satisfies $m+\kappa w=0$. Therefore, the localized state at $m>0$, $\kappa<0$ indicates the nonlinearity-induced topological phase transition associated with the amplitude-dependent Chern number. We note that the nonvanishing amplitude at the limit of $x\rightarrow \infty$ indicates that it is impossible to normalize the edge mode. Meanwhile, in finite systems, such a nonvanishing localized mode can be normalized and thus can robustly emerge. We can also check that no localized modes appear in the case of positive $m$ and $\kappa$, where the nonlinear Chern number is $C_{\rm NL} = -1/2$ at any amplitudes.

In lattice systems, we numerically validate the existence of the amplitude-dependent gapless modes and their correspondence to the nonlinear Chern number. We reveal that the finite-size effect leads to nonzero gaps of the localized modes, while such band gaps converge to zero in the thermodynamic limit $L\rightarrow \infty$. We also find that discontinuity of edge modes can alter the phase boundary, and thus lattice systems exhibit the bulk-boundary correspondence at the parameters where they well-approximate the continuum models. To show that, we consider the rediscretization of the continuum nonlinear Dirac Hamiltonian in Eq.~\eqref{nonlinearDirac} and reveal the correspondence between the parameters in the nonlinear QWZ model (Eq.~\eqref{model-wavenumber}) and the lattice constant (see Supplementary Information for the detail). Using the small lattice constant $h$, one can assume that the discretized Hamiltonian reproduces the behavior of the continuum nonlinear system.

\begin{figure*}
\includegraphics[width=140mm, bb=0 0 1410 705,clip]{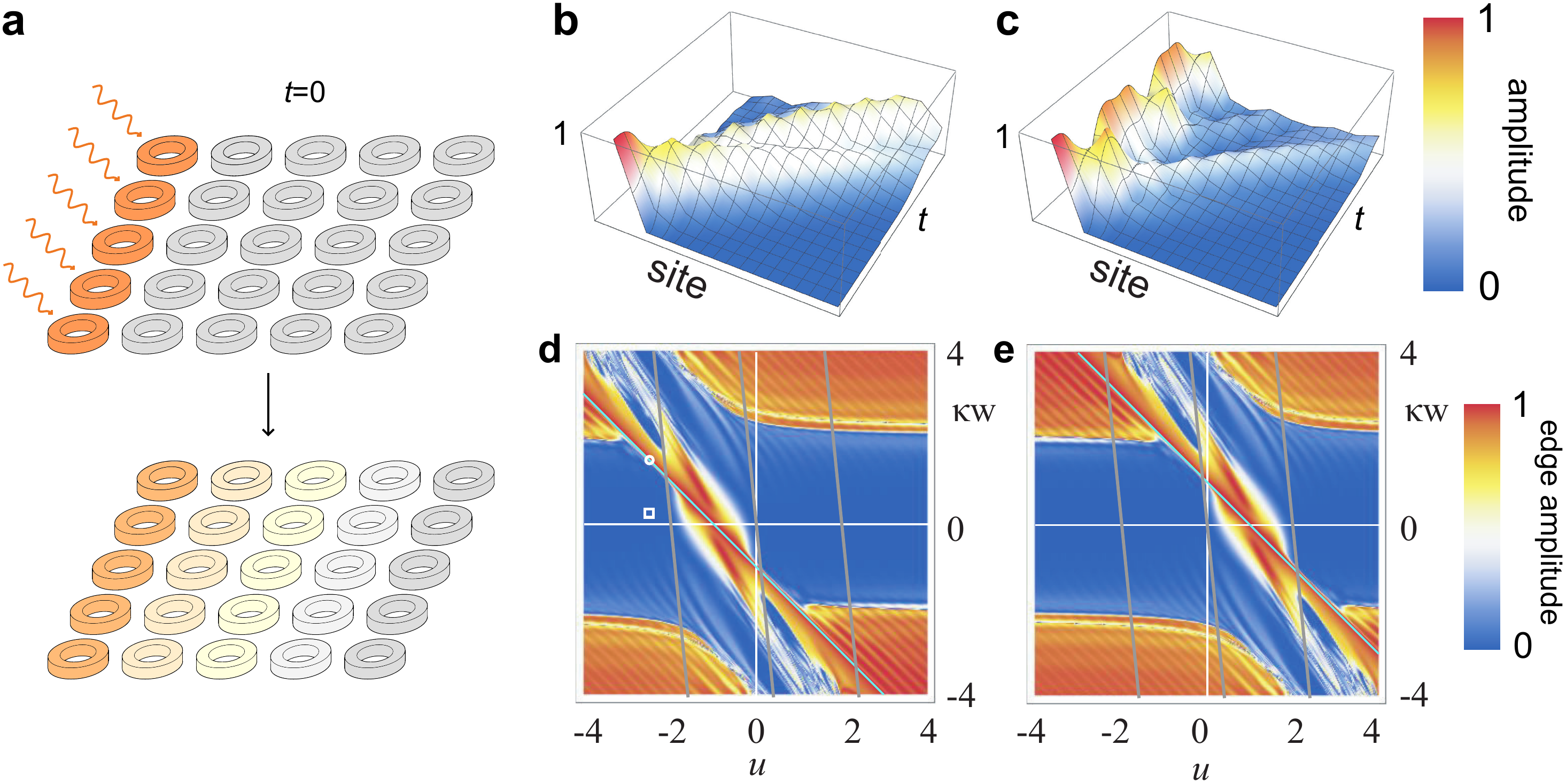}
\caption{\label{fig6} {\bf Phase diagram of the quench dynamics of the nonlinear QWZ model.} {\bf a,} The schematics of the experimental protocol of the quench dynamics is shown. First, one excites the edge sites by, e.g., applying lasers to the edge resonators in nonlinear topological photonic insulators. Then, one observes the nonlinear dynamics without external fields and confirms the existence or absence of a long-lived localized state. {\bf b}, We plot the time evolution of the quench dynamics in a trivial phase. We use the parameters $u=-2.5$, $\kappa=1.5$, and $w=1$, which correspond to the white square in panel \textbf{d}. We confirm the absence of localized edge modes. {\bf c}, We plot the time evolution of the quench dynamics of nonlinear edge modes. We use the parameters $u=-2.5$, $\kappa=0.25$, and $w=1$, which correspond to the white circle in panel \textbf{d}. We confirm that a localized state remains for a long time, which indicates the existence of edge modes. {\bf d} and {\bf e,} We simulate the quench dynamics and plot the amplitude remaining at the edge sites in the long-term limit after a quench in the $u$-$\kappa w$ plane. The initial configuration is taken as the edge modes in the linear limit at $u=-1$ ($u=1$) in panel {\bf d} ({\bf e}). The light-blue lines indicate the parameters where we can obtain exact edge-localized solutions. The gray lines represent the phase boundary derived from the nonlinear Chern number. These lines agree with the right (left) boundary of the topological phase in panel {\bf d} ({\bf e}), and thus the quench dynamics shows the shift of the phase boundary by the nonlinearity.
}
\end{figure*}

To numerically confirm the bulk-boundary correspondence in the lattice model, we calculate the minimum of the absolute values of the nonlinear eigenvalues at different sizes. We fix the parameters $m=-1$ and $\kappa w/L=1$, where the nonlinear band structure is gapless in the corresponding continuum model. Figure \ref{fig5} shows the size and lattice-constant dependencies of the energy gaps. We obtain nonzero gaps even at the topological parameters $C_{\rm NL}\neq 0$ due to the finite-size effect. However, the gap becomes smaller as the system size becomes larger. Specifically, if the lattice constant $h$ is small enough to reproduce the behavior of the continuum Hamiltonian, we confirm that the gap is proportional to $1/(L+1/2)$ with $L$ being the system size, which corresponds to the analytical solution of the eigenvalue equation of the linear QWZ model (see Supplementary Information). Thus, the gap will be closed in the continuum and thermodynamic limit. Meanwhile, we find sudden decreases in the sizes of the band gaps at $h>0.14$. The inconsistency between the numerical results at large lattice constants and the analytical estimation in the continuum limit indicates that the bulk-boundary correspondence can be modified by strong nonlinearity (see Supplementary Information for additional numerical calculations).

{\it Observation protocol of nonlinear edge modes via the quench dynamics.---}
In realistic setups, it is difficult to directly prepare a single nonlinear edge mode because it has a complicated amplitude distribution and thus one needs fine-tuning of the strength of the excitation of each site. Instead, one can observe the topological properties via quench dynamics. In quench dynamics, one only has to excite the edge sites at homogeneous amplitudes and observe the dynamics without any other external interactions as depicted in Fig.~\ref{fig6}a. 

To confirm the correspondence between the existence of the nonlinear edge modes and the localized states in quench dynamics, we numerically simulate quench dynamics of the nonlinear QWZ model (Eq.~\eqref{model-wavenumber}) at various parameters. Figures \ref{fig6}b and c show the time evolution of the quench dynamics with and without nonlinear edge modes. We also obtain the phase diagrams in  Figs.~\ref{fig6}d and e, which are classified by the amplitude at the edge of the sample in the long-time limit. Here, we consider two initial states equivalent to the nonlinear edge modes at $u+\kappa w= \pm 1$ (see also Supplementary Information). In weakly nonlinear regimes, the localized states remain in the topological cases and vanishes in the trivial cases. We can also confirm the shift of the phase boundary by the change of the amplitude, which indicates the existence of the nonlinearity-induced topological phase transitions. We note that strong nonlinearity induces trap phases \cite{Eilbeck1985,Ezawa2021,Ezawa2022} where the localized initial states remain without topological origins (see also Supplementary Information). It is also noteworthy that the phase boundary obtained from quench dynamics does not exactly match that of the nonlinear Chern number because the perfectly-localized initial condition is quite different from the Bloch wave and thus quench dynamics cannot fully reproduce the topological properties obtained from the Bloch ansatz.

{\it Possible experimental setups of nonlinear Chern insulators.---}
In our model, we considered the Kerr-like nonlinearity, which is common in nonlinear optics as well as the mean-field analysis of interacting bosons \cite{Boyd2003,Gross1961,Pitaevskii1961}. Thus, a nonlinear Chern insulator can be realized in various classical and quantum systems. One possible concern is the nonlinear cross terms in the nonlinear QWZ model $|\Psi_1|^2\Psi_2$ and $|\Psi_2|^2\Psi_1$, which may be unfeasible in realistic setups. However, we can omit such cross terms in nonlinear Chern insulators and instead use on-site Kerr nonlinear terms, $\kappa |\Psi_j(x,y)|^2 \Psi_j(x,y)$ (see also Supplementary Information). Introducing such on-site Kerr nonlinearity, one can realize nonlinear Chern insulators by using the near-term techniques of topological and nonlinear physics.

Specifically, previous studies have argued topological photonics using ring-resonator arrays \cite{Hafezi2011,Mittal2019}. By focusing on a pseudo-spin sector in light propagating through ring-resonator arrays, one can realize the photonic counterpart of a Chern insulator. If we utilize different materials in the ring resonators at different sublattices, one can possibly tune the sign of the Kerr nonlinear effect \cite{Swartzlander1991} at each site. Thus, topological photonic metamaterials using different resonators can be a candidate system to realize the nonlinear topological insulators analyzed in the present paper.

{\it Discussion.---}
We introduced the nonlinear Chern number in two-dimensional systems by considering the nonlinear eigenvalue problem (Eq.~\eqref{NLeigen}). We theoretically proved the bulk-boundary correspondence of the nonlinear Chern number in weakly nonlinear regimes, which guarantees that if the nonlinearity is small enough compared to the bulk band gap, the bulk topology can predict the existence of the edge modes. More importantly, we investigated the bulk-boundary correspondence in the stronger nonlinear regime where the nonlinearity is larger than the linear band gap while it is smaller than the linear couplings. We showed the existence of the nonlinearity-induced topological phase transition that depends on the amplitude and thus has no counterparts in linear systems. We proposed a minimal model of a nonlinear Chern insulator, whose exact bulk solutions indicate the existence of the nonlinearity-induced topological phase transition detected by the nonlinear Chern number. We analytically show that the nonlinear Chern number exactly predicts the nonlinearity-induced topological phase transition in the nonlinear Dirac Hamiltonian, implying the nonlinear bulk-boundary correspondence.. In a nonlinear lattice system, we numerically checked that the bulk-boundary correspondence is recovered in the continuum and thermodynamic limit.

Our results indicate the existence of unique topological phenomena beyond the weakly nonlinear regime. There remain intriguing issues to fully establish the topological classification of nonlinear systems including further strongly nonlinear cases where nonlinearity is even stronger than the linear couplings. Since such strong nonlinearity can induce bulk-localized modes \cite{Eilbeck1985,Lumer2013,Ezawa2021,Ezawa2022} that are out of the description of the Bloch ansatz, it is unclear whether or not the nonlinear Chern number still fully works. 

From the perspective of many-body quantum physics, nonlinear systems are regarded as the result of the mean-field approximation of interacting many-body systems \cite{Gross1961,Pitaevskii1961}. The nonlinear Chern number should correspond to the topology of the excited states of interacting systems, which we can observe by applying external fields oscillating at the corresponding frequency. In fact, the nonlinear eigenvalues correspond to the energies per particle of the many-body wavefunctions in the mean-field analysis (see Supplementary Information), which are not the ground state except for that of the minimum eigenvalue. Therefore, the nonlinear topology would reveal the unexplored topological phases in quantum many-body systems at the level of mean-field approximation.

We thank Zongping Gong, Takahiro Morimoto, Taro Sawada, and Haruki Watanabe for valuable discussions. K.S. is supported by World-leading Innovative Graduate Study Program for Materials Research, Information, and Technology (MERIT-WINGS) of the University of Tokyo. K.S. is also supported by JSPS KAKENHI Grant Number JP21J20199. M.E. is supported by JST, CREST Grants Number JPMJCR20T2 and Grants-in-Aid for Scientific Research from MEXT KAKENHI (Grant No. 23H00171). Y.A. acknowledges support from the Japan Society for the Promotion of Science through Grant No. JP19K23424 and from JST FOREST Program (Grant Number JPMJFR222U, Japan). N.Y. acknowledges support from the Japan Science and Technology Agency (JST) PRESTO under Grant No. JPMJPR2119 and JST Grant No. JPMJPF2221. T.S. is supported by JSPS KAKENHI Grant Numbers JP19H05796, JST, CREST Grant Number JPMJCR20C1, and the JST ERATO-FS Grant No. JPMJER2204. N.Y. and T.S. are also supported by Institute of AI and Beyond of the University of Tokyo.

\widetext
\pagebreak
\begin{center}
\textbf{\large Supplementary Materials}
\end{center}

\renewcommand{\theequation}{S\arabic{equation}}
\renewcommand{\thefigure}{S\arabic{figure}}
\setcounter{equation}{0}
\setcounter{figure}{0}

\subsection{Justification of the Bloch ansatz via the perturbation analysis.}
If the nonlinear term is small compared to the linear band gap, one can regard the nonlinear effect as a perturbation to the linear band structure. Under such an assumption, one can perturbatively calculate the nonlinear eigenvectors. To show the perturbation-calculation protocol of the nonlinear eigenvalue problem we rewrite the nonlinear eigenequation as 
\begin{equation}
H(\Psi) \Psi = (H^0 + \kappa H_{\rm NL}(\Psi)) \Psi = E\Psi. \label{mod-nonlinear-eigprob}
\end{equation}
We consider the perturbation expansion by $\kappa$,
\begin{eqnarray}
\Psi &=& \Psi^{(0)} + \kappa \Psi^{(1)} + \cdots, \label{perturb_eigvec}\\
E &=& E^{(0)} + \kappa E^{(1)} + \cdots. \label{perturb_eigval}
\end{eqnarray}
One can determine $\Psi^{(0)}$ and $E^{(0)}$ from the eigenvalue and eigenvector of the linear Hamiltonian $H^0$
\begin{equation}
H^0 \Psi^{(0)} = E\Psi^{(0)}.
\end{equation}
Then, the first-order perturbation is calculated from the eigenequation of $(H^0 + \kappa H_{\rm NL}(\psi^{(0)}))$ as 
\begin{equation}
(H^0 + \kappa H_{\rm NL}(\Psi^{(0)})) (\Psi^{(0)}+\kappa \Psi^{(1)}) = (E^{(0)}+\kappa E^{(1)}) (\Psi^{(0)}+\kappa \Psi^{(1)}).\label{first-order-perturb}
\end{equation}
One can confirm the consistency between Eqs.~\eqref{mod-nonlinear-eigprob} and \eqref{first-order-perturb} by substituting \eqref{perturb_eigvec} and \eqref{perturb_eigval} into the former equation.

In translation-invariant systems, the nonperturbed eigenvector $\Psi^{(0)}$ is described by a Bloch wave $\Psi^{(0)} = e^{i\mathbf{k}\mathbf{x}}\psi^{(0)}$, due to the Bloch theorem of linear systems. Since the Bloch wave exhibits no site-dependence of the amplitude, one can assume that the nonlinear term $\kappa H_{\rm NL}(\Psi)$ is also uniform, and thus the whole effective Hamiltonian $H^0 + \kappa H_{\rm NL}(\Psi)$ still has the translational symmetry. Therefore, in weakly nonlinear systems, one can believe that all of the nonlinear eigenvectors are described by the Bloch ansatz. We note that in strongly nonlinear regimes, there can be localized modes that cannot be described by the Bloch ansatz \cite{Eilbeck1985,Ezawa2021,Ezawa2022}. However, the periodic solutions obtained from the Bloch ansatz are still exact nonlinear eigenvectors under the periodic boundary conditions.

\subsection{Real-space description of the nonlinear QWZ model.} To investigate the existence or absence of topological edge modes in lattice systems, we construct a minimal lattice model of a nonlinear Chern insulator, which we term the nonlinear QWZ model (see Eq.~\eqref{model-wavenumber} for the wavenumber-space description). Its real-space dynamics is described as
\begin{eqnarray}
i\frac{d}{dt}\Psi_j(x,y) &=& \sum_l \{(\sigma_z)_{jl} [2u\Psi_l(x,y)+\Psi_l(x+1,y)+\Psi_l(x-1,y) +\Psi_l(x,y+1)+\Psi_l(x,y-1)]/2 \nonumber\\
&{}& + (\sigma_x)_{jl}(\Psi_l(x+1,y)-\Psi_l(x-1,y))/2i + (\sigma_y)_{jl}(\Psi_l(x,y+1)-\Psi_l(x,y-1))/2i\} \nonumber\\
&{}& - (-1)^j \kappa (|\Psi_1(x,y)|^2+|\Psi_2(x,y)|^2) \Psi_j(x,y),\label{model-real}
\end{eqnarray}
where $j$, $l$ and $x$, $y$ represent the internal degree of freedom and the location, respectively, and $\Psi_j(x,y)$ is the $j$-th component of the state vector at the location $(x,y)$. $(\sigma_i)_{jl}$'s are the $(j,l)$-component of the $i$th Pauli matrices. This lattice model introduces the staggered Kerr-like nonlinearity $- (-1)^j \kappa (|\Psi_1(x,y)|^2+|\Psi_2(x,y)|^2) \Psi_i(x,y)$ to the linear QWZ model \cite{Qi2006}.

\subsection{Exact bulk solutions of the nonlinear QWZ model.} To obtain the phase diagram in Fig.~\ref{fig2}b, we analytically solve the nonlinear eigenequation in Eq.~\eqref{model-wavenumber}. If we focus on special solutions where the squared amplitude $|\Psi_1(\mathbf{k})|^2+|\Psi_2(\mathbf{k})|^2=w$ has no $\mathbf{k}$-dependence, the nonlinear eigenequation (Eq.~\eqref{model-wavenumber}) exactly corresponds to a linear one. Therefore, by solving the corresponding linear eigenequation, we obtain the following exact bulk eigenvalues and eigenvectors,
\begin{eqnarray}
E_{\pm }\left( k_{x},k_{y}\right) &=& \pm \sqrt{2\cos k_{x}\cos k_{y}
+2\left(u+\kappa w\right) \left( \cos k_{x}+\cos k_{y}\right) +\left( u+\kappa w\right) ^{2}+2},\label{exact_eigval}\\
\left( 
\begin{array}{c}
\psi _{1\pm}\left( \mathbf{k}\right)  \\ 
\psi _{2\pm}\left( \mathbf{k}\right) 
\end{array}
\right) &=&  \frac{\sqrt{w}}{ c_{\pm}\left( \mathbf{k}\right)} \left( 
\begin{array}{c}
u+\kappa w+\cos k_{x}+\cos k_{y}+E_{\pm }\left( k_{x},k_{y}\right)  \\ 
\sin k_{x}-i\sin k_{y}
\end{array}
\right),\label{exact_eigvec}
\end{eqnarray}
where $c_{\pm}\left( \mathbf{k}\right)=\sqrt{\left( u+\kappa w+\cos k_{x}+\cos k_{y}+E_{\pm} \left( k_{x},k_{y}\right) \right) ^{2}+\sin^{2}k_{x}+\sin ^{2}k_{y}}$ is a normalization constant. By using these nonlinear eigenvectors, we analytically obtain the nonlinear Chern number,
\begin{eqnarray}
C_{\rm NL} = \begin{cases}
1 & (0 < u+\kappa w< 2) \\
-1 & (-2 < u+\kappa w< 0) \\
0 & ({\rm otherwise})
\end{cases},\label{lattice-chern}
\end{eqnarray}
as summarized in Fig.~\ref{fig2}b. We note that one can generally obtain exact bulk solutions if the nonlinear equation has the form in Eq.~\eqref{general_nonlinear_eq} (see also the following section).

\subsection{Derivation of exact bulk solutions.}
If the nonlinear dynamics is described as 
\begin{equation}
 i\partial _{t} \psi _{n}\left( \mathbf{k};t\right)
=\sum_{m}f_{nm}\left( \sum_{j=1}^{N}\left\vert \psi _{j}\left( \mathbf{k};t\right) \right\vert ^{2}\right) \psi _{m}\left(\mathbf{k};t\right), \label{Eq1}
\end{equation}
one can obtain its exact bulk nonlinear eigenvectors. Here, $\psi _{n}\left( \mathbf{k}\right) $ is the state variable in the momentum space $\mathbf{k}$, and $n$, $m$ and $j$ label the unit cell or the internal degree of freedom such as the spin satisfying $1\leq n\leq N$, $1\leq m\leq N$ and $1\leq j\leq N$. 

The basic idea is to construct a set of special exact solutions self-consistently by requiring that the quantity $w$ defined by 
\begin{equation}
w=\sum_{j=1}^{N}\left\vert \psi _{j}\left( \mathbf{k}\right) \right\vert^{2}
\label{wParam}
\end{equation}
has no $\mathbf{k}$ dependence. Then, the nonlinear eigenequation becomes a linear equation 
\begin{equation}
E \psi _{n}\left( \mathbf{k}\right)
=\sum_{m}f_{nm}\left( w\right) \psi _{m}\left( \mathbf{k}\right), \label{Fw}
\end{equation}
which is diagonalizable by using eigenfunctions $\psi _{m}^{(\alpha )}\left(\mathbf{k}\right) $, 
\begin{equation}
\sum_{m}f_{nm}\left( w\right) \psi _{m}^{\left( \alpha \right) }\left(\mathbf{k}\right) =E^{\left( \alpha \right) }\psi _{n}^{\left( \alpha\right)}\left( \mathbf{k}\right)
\end{equation}
together with the eigenenergy $E^{\left( \alpha \right) }$. We note that 
\begin{equation}
\tilde{\psi}_{m}^{\left( \alpha \right) }\left( \mathbf{k}\right) \equiv c\left( \mathbf{k}\right) \psi _{m}^{\left( \alpha \right) }\left( \mathbf{k}\right)  \label{psi}
\end{equation}
is also an eigenfunction of the same eigenequation in the linear case, where $c\left( \mathbf{k}\right) $ is an arbitrary function. When we set  $c\left( \mathbf{k}\right) $ independently of $\mathbf{k}$ as
\begin{equation}
c\left( \mathbf{k}\right) \equiv \frac{\sqrt{w}}{\sqrt{\sum_{j=1}^{N}\left \vert \psi _{j}^{\left( \alpha \right) }\left( \mathbf{k}\right) \right\vert^{2}}},
\end{equation}
we obtain 
\begin{equation}
\sum_{j=1}^{N}\left\vert \tilde{\psi}_{j}\left( \mathbf{k}\right) \right\vert ^{2}=w.  \label{SumW}
\end{equation}
It follows from Supplementary Eqs.~\eqref{Fw} and \eqref{SumW} that 
\begin{equation}
E \tilde{\psi}_{n}^{\left( \alpha \right) }\left(
\mathbf{k}\right) =\sum_{m}f_{nm}\left( \sum_{j=1}^{N}\left\vert \tilde{\psi}_{j}^{\left( \alpha \right) } \left( \mathbf{k}\right)
\right\vert^{2}\right) \tilde{\psi}_{m}^{\left( \alpha \right) }\left( 
\mathbf{k}\right) ,
\end{equation}
which indicates that Supplementary Eq.~\eqref{psi} is the nonlinear eigenvector of Supplementary Eq.~\eqref{Fw}. Hence, we have solved Supplementary Eq.~\eqref{Fw} self-consistently by requiring Supplementary Eq.~\eqref{wParam}. Supplementary Equation \eqref{psi} presents a set of exact solutions parametrized by the real number $w$ in terms of solutions of the linear equation (\ref{Fw}).

\subsection{Perturbation analysis of the bulk modes of the nonlinear QWZ model.} While we calculate the exact solutions of the bulk modes of the nonlinear QWZ model in the main text, we can also obtain the same bulk modes from the perturbation analysis or the self-consistent calculations. Specifically, if we conduct the perturbation analysis described in the previous section, the calculation stops at the first-order perturbation and derives the same bulk modes as the exact solutions.

The zeroth-order solutions, i.e., the linear solutions of the bulk modes of the QWZ model are 
\begin{eqnarray}
\left( 
\begin{array}{c}
\psi^{(0)} _{1\pm}\left( \mathbf{k}\right)  \\ 
\psi^{(0)} _{2\pm}\left( \mathbf{k}\right) 
\end{array}
\right) &=&  \frac{\sqrt{w}}{ c_{\pm}\left( \mathbf{k}\right)} \left( 
\begin{array}{c}
u+\cos k_{x}+\cos k_{y}+E_{\pm }\left( k_{x},k_{y}\right)  \\ 
\sin k_{x}-i\sin k_{y}
\end{array}
\right),\\
E_{\pm }\left( k_{x},k_{y}\right) &=& \pm \sqrt{2\cos k_{x}\cos k_{y}
+2u \left( \cos k_{x}+\cos k_{y}\right) +u^2+2}, \\
c_{\pm}\left( \mathbf{k}\right) &=& \sqrt{\left( u+\cos k_{x}+\cos k_{y}+E_{\pm} \left( k_{x},k_{y}\right) \right) ^{2}+\sin^{2}k_{x}+\sin ^{2}k_{y}},
\end{eqnarray}
where we fix the norm as $|\psi _{1\pm}\left( \mathbf{k}\right) |^2+ |\psi _{2\pm}\left( \mathbf{k}\right)|^2=w$. Then, substituting these solutions into the state-dependent Hamiltonian of the nonlinear QWZ model, one can obtain the first-order-perturbation solutions. Due to the nonlinear terms only depending on the norm of the nonlinear eigenvector, the substituted effective Hamiltonian is described as 
\begin{eqnarray}
H(\mathbf{k},\psi(\mathbf{k})) = \left(
  \begin{array}{cc}
   u + \kappa w + \cos k_x+\cos k_y  & \sin k_x + i\sin k_y \\
   \sin k_x - i\sin k_y & -(u + \kappa w +\cos k_x+\cos k_y)
  \end{array}
  \right), \label{model-wavenumber-perturbcalc}
\end{eqnarray}
independently of the wavenumber. The eigenvalues and eigenvectors of this Hamiltonian are the same as those of the nonlinear QWZ model (Eqs.~\eqref{exact_eigval} and \eqref{exact_eigvec}), and thus the first-order perturbation calculation is consistent with the exact solutions.

The self-consistent calculation is equivalent to the higher-order perturbation calculation (see also the following section). However, if one substitutes the first-order solutions into the state-dependent Hamiltonian of the nonlinear QWZ model, one obtains the same effective Hamiltonian as Eq.~\eqref{model-wavenumber-perturbcalc}. Therefore, the nonlinear eigenvectors and eigenvalues obtained from the self-consistent calculation are the same as those obtained from the first-order perturbation and the exact solutions.

\subsection{Numerical simulations of the real-space dynamics of the nonlinear QWZ model.} In Fig.~\ref{fig2}, we numerically calculate the dynamics of the nonlinear QWZ model by using the fourth-order Runge-Kutta method. We consider the $20\times20$ square lattice, where each lattice point has two internal degrees of freedom. We impose the open boundary condition in the $x$ direction and the periodic boundary condition in the $y$ direction. We set the time step $dt = 0.005$. The simulation starts from the localized initial states where  We use the parameters $u=3$, $\kappa=0.1$, $w=1$ in Fig.~\ref{fig2}c and $u=-1$, $\kappa=0.1$, $w=1$ in Fig.~\ref{fig2}d.  In these figures, we plot the square root of the sum of the square of absolute values of the first and second components.

\subsection{Self-consistent calculation of nonlinear band structures.} To obtain the nonlinear band structures in Fig.~\ref{fig3}, we numerically calculate the nonlinear eigenvalue problem by using the self-consistent method. We first rewrite the nonlinear dynamics as $i\partial_t \Psi = f(\Psi) = H(\Psi)\Psi$, where $H(\Psi)$ is a state-dependent effective Hamiltonian. Then, we conduct the self-consistent calculation in the following procedure: (i) We numerically diagonalize $H(\vec{0})$ and set the initial guess of the eigenvalue and eigenvector $\Psi_0$, $E_0$ by adopting a pair of the obtained eigenvalue and eigenvector of $H(\vec{0})$. We fix the norm of $\Psi_0$ to be $||\Psi_0||^2=w$. (ii) We substitute the guessed eigenvector $\Psi_i$ after $i$ iterations into $H(\Psi)$ and diagonalize $H(\Psi_i)$. (iii) We choose the obtained eigenvalue that is the closest to the previous guess $E_i$, and the corresponding eigenvector as the next guess $E_{i+1}$, $\Psi_{i+1}$. (iv) We iterate the steps (ii) and (iii) until the distance between $\Psi_{i}$ and $\Psi_{i+1}$ becomes smaller than the threshold, $||\Psi_{i+1} - \Psi_{i}|| < \epsilon$, or the iteration reaches a fixed number. We also perform these calculations starting from all the eigenvectors of $H(\vec{0})$ and obtain a set of nonlinear eigenvectors and eigenvalues of $f(\Psi)$.

In Fig.~\ref{fig3}, we consider the parameterized state-dependent Hamiltonian that corresponds to the nonlinear Chern insulator under the assumptions of the $y$-periodic Bloch ansatz and the open boundary condition in the $x$ direction. The state-dependent Hamiltonian is described as
\begin{eqnarray}
H(\Psi) = \left(
  \begin{array}{cc}
   u + \Delta_x^2 + \cos k_y + \kappa (|\Psi_1|^2+|\Psi_2|^2) & (i\Delta_x + i\sin k_y) \\
   (i\Delta_x -i\sin k_y) & -(u + \Delta_x^2 + \cos k_y + \kappa (|\Psi_1|^2+|\Psi_2|^2))
  \end{array}
  \right),
\end{eqnarray}
where $\Delta_x$ and $\Delta_x^2$ are the difference operators defined as $\Delta_x \Psi(x)=[\Psi(x+1)-\Psi(x)]/2$ and $\Delta_x^2 \Psi(x)=\Psi(x+1)+\Psi(x-1)-2\Psi(x)$, respectively. Then, we calculate the nonlinear eigenvalues of this state-dependent Hamiltonian at $k_y=n\Delta k$, $n=-N,-N+1,\cdots,N-1,N$, and $\Delta k=\pi/N$ with $N$ being $N=50$.

To calculate the nonlinear band gap in Fig.~\ref{fig5}, we also use the self-consistent method. However, to stably obtain the band structure of topological gapless modes, we start from the eigenvectors of $H((\sqrt{w/2L},\sqrt{w/2L},\cdots)^T)$ instead of those of $H(\vec{0})$, where $L$ is the system size. This is because $H(\vec{0})$ has no topological gapless modes, and thus the initial guess corresponding to the gapless modes in $H(\Psi)$ cannot be obtained from $H(\vec{0})$. We also note that the self-consistent calculation is only stable at weak nonlinearity. To calculate the band structures in strongly nonlinear systems than those analyzed in the main text, one should instead use the Newton method, whose details are described in the following section.

\subsection{Quasi-Newton method to solve strongly nonlinear eigenvalue problems.}
Here, we explain the (quasi-)Newton method to solve nonlinear eigenvalue problems in strongly nonlinear regimes. While we use the self-consistent calculation in weakly nonlinear systems, it often fails to converge in strongly nonlinear regimes. Instead, we should use the (quasi-)Newton method \cite{Fletcher1987}, which approximately calculates the roots of algebraic equations. To apply the Newton method, we reconsider the set of the nonlinear eigenvalue problem and the normalization condition,
\begin{eqnarray}
 f(\Psi) &=& E\Psi, \\
 ||\Psi||_2^2 &=& w.
\end{eqnarray}
If we consider the $n$-component vector $\Psi$, there are $n+1$ unknown variables consisting of the set of the components of the eigenvector $\Psi_i$ and the eigenvalue $E$. Then, the nonlinear eigenequation provides $n$ algebraic equations and the normalization condition does one algebraic equation. Therefore, the total number of algebraic equations is equal to the number of unknown variables, and thus we can solve the eigenvalue problem by calculating the roots of the algebraic equations. 

The Newton method solves the algebraic equations by using the information of the residual and the local gradient. We rewrite the equations as $F(\Psi,E)=0$, where $F_i$ is the $i$th component of $f(\Psi) - E\Psi$ for $i=1,\cdots,n$ and $F_{n+1}$ is $F_{n+1}(\Psi,E) = ||\Psi||_2^2 - w$. We first set an initial guess of the eigenvalue $E_0$ and eigenvector $\Psi_0$. In our calculations, we randomly determine the initial guess of the eigenvalue from the uniform distribution $[-2,2]$ and the real part and the imaginary part of each component of the initial guess of the eigenvector from the uniform distribution $[-1,1]$. Then, we renormalize the initial guess of the eigenvector. If the guess is close to the genuine solution $\Psi_g$, $E_g$, we can approximate the residual as 
\begin{eqnarray}
 F(\Psi_0,E_0) = J_F (\Psi_0-\Psi_g, E_0-E_g) (\Psi_0-\Psi_g, E_0-E_g)^T + (\text{higher-order terms}),
\end{eqnarray}
where $J_F$ is the Jacobian matrix of $F$. Thus, $(\Psi_1,E_1) = (\Psi_0,E_0) + \alpha J_F^{-1}F(\Psi_0,E_0)$ with $\alpha$ being the update step becomes better approximations of the nonlinear eigenvector and eigenvalue. Iterating this update procedure of the approximated eigenvector and eigenvalue, 
\begin{eqnarray}
 (\Psi_{i+1},E_{i+1}) = (\Psi_i,E_i) + \alpha J_F^{-1}F(\Psi_i,E_i), \label{quasi-newton}
\end{eqnarray}
we can numerically calculate the nonlinear eigenvector and eigenvalue. In our calculations, we fixed the update step as $\alpha=0.01$. If we start from different initial guesses, we can obtain other sets of eigenvectors and eigenvalues, which can cover all the eigenvectors of the nonlinear equation. 

Since the Jacobian matrix of $F$ is tough to calculate in our model, we also numerically obtain the approximation of the Jacobi matrix, which is known as the quasi-Newton method. Specifically, we use an update algorithm called the Broyden method \cite{Broyden1965}. We denote the approximated Jacobian matrix at the $i$th step of the quasi-Newton method by $B_i$. We set $B_0$ as the identity matrix, $B_0=I$. Then, the approximated Jacobian matrix is updated as 
\begin{equation}
 B_{i+1} = B_i + \frac{(y_i-B_i \Delta \Psi'_i) (\Delta \Psi'_i)^T}{(\Delta \Psi'_i)^T \Delta \Psi'_i},
\end{equation}
where $y_i$ and $\Delta \Psi'_i$ are $y_i=F(\Psi_{i+1},E_{i+1}) - F(\Psi_i,E_i)$ and $\Delta \Psi'_i = (\Psi_{i+1}-\Psi_i,E_{i+1}-E_i)$.
Finally, we substitute $B_i$ into $J_F$ in Supplementary Eq.~\eqref{quasi-newton} and update the guess of eigenvectors and eigenvalues.

\subsection{Theorem of the bulk-boundary correspondence in weakly nonlinear systems.} We mathematically show the bulk-boundary correspondence in weakly nonlinear systems. We here use a simple notation $f(\vec{\Psi}) = E\vec{\Psi}$ instead of $f_j(\Psi;\mathbf{r}) = E\Psi_j (\mathbf{r})$, where $\vec{\Psi}$ is the nonlinear eigenvector whose components correspond to the state variables at the locations $\mathbf{r}$ and the internal degrees of freedom $j$. The claim of the theorem is as follows: 

{\it Theorem:} Suppose $f(\vec{\Psi}) = E\vec{\Psi}$ is a nonlinear eigenvalue problem on a two-dimensional lattice system that satisfies the following assumptions: (1) When we rewrite the nonlinear function $f$ as $f(\vec{\Psi})=H(\vec{\Psi})\vec{\Psi}$, there exists a positive real number $c<1$ that satisfies $||H(\vec{\Psi})-H(\vec{0})|| < gc/2$ for any complex vector $\vec{\Psi}$ with $g$ being the bandgap of $H(\vec{0})$ and $||\cdot||$ being the operator norm. (2) There exists a positive real number $c'<1$ such that for any pairs of complex vectors $\Psi$ and $\Psi+\Delta\Psi$ with the norm $w$, they satisfy $||H(\vec{\Psi}+\Delta\vec{\Psi})-H(\vec{\Psi})|| \leq g(1-c)c'(6\sqrt{2}w)^{-1} ||\Delta\vec{\Psi}||$. (3) For any complex vector $\vec{\Psi}$, one can rewrite the nonlinear function $f(\vec{\Psi})$ as $f(\vec{\Psi})=\tilde{H}(\vec{\Psi})\vec{\Psi}$, where $\tilde{H}$ is a Hermitian matrix. (4) The nonlinear equation satisfies the $U(1)$ symmetry, $f(e^{i\theta}\vec{\Psi})=e^{i\theta}f(\vec{\Psi})$. We also assume that the number of nonlinear eigenvectors is equal to that of the linear eigenvectors of $H(\vec{0})$. Then, the nonlinear eigenequation $f(\vec{\Psi}) = E\vec{\Psi}$ exhibits robust gapless boundary modes if and only if its nonlinear Chern number is nonzero.

To prove Theorem, we first show the following proposition:

{\it Proposition 1:} In the nonlinear eigenvalue problem that satisfies the assumptions in Theorem, the self-consistent calculation converges, $E_i \rightarrow E_{\infty}$, $\vec{\Psi}_i \rightarrow \vec{\Psi}_{\infty}$. Furthermore, there exists an eigenvector $\vec{\Psi}_0$ and eigenvalue $E_0$ of $H(\vec{0})$ that satisfy $||E_{\infty}-E_0|| < gc/2(1-c')$ and $||\vec{\Psi}_{\infty}-\vec{\Psi}_0|| < 2^{-1/2}c(1-2c')^{-1}w$.

To show this proposition, we utilize the perturbation theorem of the eigenvectors in linear systems: Suppose $H$ as a nondegenerate Hermitian matrix and $g$ as the minimum difference between its two eigenvalues. If $||A||<g/2$ in terms of the operator norm, for an arbitrary eigenvector $\vec{\Psi}$ of $H+A$, there exists an eigenvector $\vec{\Psi}_0$ of $H$ that satisfies 
$||\vec{\Psi}-\vec{\Psi}_0|| \leq D g^{-1}||A||\,||\vec{\Psi}_0||$ ($D=2\sqrt{2}$). We iteratively use this theorem and evaluate the distance between the guess of eigenvectors $\vec{\Psi}_i$ and $\vec{\Psi}_{i+1}$ at each step. 

We also show that the resulting eigenvalue and eigenvector, $E_{\infty}$ and $\vec{\Psi}_{\infty}$ are indeed a pair of nonlinear eigenvalue and eigenvector of $f(\vec{\Psi})$, which is summarized in 

{\it Proposition 2:} The converged solutions of the self-consistent calculation, $E_{\infty}$ and $\vec{\Psi}_{\infty}$ satisfy $f(\vec{\Psi}_{\infty}) = E_{\infty}\vec{\Psi}_{\infty}$. 

We show this proposition by using simple inequalities and limit evaluations. By using these propositions, we finally show the following lemma to prove Theorem. 

{\it Lemma:} For an arbitrary eigenvector of a nonlinear eigenvalue problem $(H+f(\vec{\Psi})) \vec{\Psi}=E\vec{\Psi}$ that satisfies the conditions in Theorem, there exists the eigenvector of $[H+(1-\epsilon)f(\vec{\Psi}_{\epsilon})] \vec{\Psi}_{\epsilon}=E\vec{\Psi}_{\epsilon}$ that satisfies $||\vec{\Psi}-\vec{\Psi}_{\epsilon}||<Cw\epsilon$ ($0<\epsilon<g'/g$ with $g'$ being the minimum difference of the eigenvalues of $H+f(\vec{\Psi})$), where $w$ is the norm of $\vec{\Psi}$ and $\vec{\Psi}_{\epsilon}$, and $C$ is the constant independent of the eigenvector $\vec{\Psi}$ and the constant $\epsilon$.

The lemma indicates that in weakly nonlinear systems, one can map the nonlinear eigenvalues and eigenvectors onto those of a linear eigenvalue problem. Thus, we can show the bulk-boundary correspondence in weakly nonlinear systems.

\subsection{Proof of the main theorem.}
In this section, we prove Theorem in the main text. As discussed in the previous section, we use two propositions and a lemma to prove the theorem. In Propositions 1 and 2, we show that the nonlinear eigenvectors are obtained as the convergent values of the self-consistent calculation. We also show Lemma by using a similar technique in Proposition 1. Lemma indicates the existence of a continuous path that connects nonlinear and linear eigenvectors. Therefore, one can conclude the bulk-boundary correspondence in weakly nonlinear systems via that in linear systems. In this section, we use $F(\vec{\Psi})=H(\vec{\Psi})\vec{\Psi}=E\vec{\Psi}$ to denote the nonlinear eigenequation, where $E$ and $\vec{\Psi}$ are the nonlinear eigenvalue and eigenvector, and $H(\vec{\Psi})$ represents the effective Hamiltonian that depends on the nonlinear eigenvector.

\subsubsection{Proof of the perturbation theorem of linear eigenvectors.}
We utilize the perturbation theorem of linear eigenvectors to prove Proposition 1. The perturbation theorem is as follows: Suppose $H$ as a nondegenerate Hermitian matrix and $g$ as the minimum difference between its two eigenvalues. If $||A||_{\infty}<g/2$ in terms of the operator norm, for an arbitrary eigenvector $\vec{\Psi}$ of $H+A$, there exists an eigenvector $\vec{\Psi}_0$ of $H$ that satisfies $||\vec{\Psi}_0||=||\vec{\Psi}||$ and $||\vec{\Psi}-\vec{\Psi}_0|| \leq D g^{-1}||A||_{\infty}||\vec{\Psi}_0||$ ($D=2\sqrt{2}$).

The perturbation theorem is proved as follows. Let $\Psi$ and $E$ be the eigenvector and eigenvalue of $H+A$. From Weyl's perturbation theorem \cite{Bhatia1997}, there exists only one pair of an eigenvector and an eigenvalue, $\vec{\Psi}'_0$, $E_0$ that satisfies $|E-E_0| \leq ||A||_{\infty} < g/2$. Then, comparing eigenequations $(H+A)\vec{\Psi}=E\vec{\Psi}$ and $H\vec{\Psi}'_0=E_0\vec{\Psi}'_0$ we obtain
\begin{equation}
(H-E_0)\Delta'\vec{\Psi} = (\Delta E-A) \vec{\Psi},
\end{equation}
where we rewrite the difference of eigenvectors and eigenvalues as $\Delta' \vec{\Psi} = \vec{\Psi} - \vec{\Psi}'_0$ and $\Delta E=E-E_0$, respectively. Then, we separate $\Delta'\vec{\Psi}$ into the components parallel and perpendicular to $\vec{\Psi}'_0$ as $\Delta'\vec{\Psi} = c\vec{\Psi}'_0+\Delta\vec{\Psi}$ with $c$ being a complex coefficient. Since we have $(H-E_0)\Delta'\vec{\Psi} = (H-E_0) \Delta\vec{\Psi}$, we can rewrite $\Delta\vec{\Psi}$ by 
\begin{equation}
\Delta\vec{\Psi} = (H-E_0)^- (\Delta E-A) \vec{\Psi}
\end{equation}
with $(H-E_0)^-$ being the generalized inverse matrix of $H-E_0$. The operator norms of $\Delta E-A$ and $(H-E_0)^-$ are bounded by $||(\Delta E-A)||_{\infty} \leq 2||A||$ and $||(H-E_0)^-||_{\infty} \leq g^{-1}$. Therefore, we obtain the bound of the norm of $\Delta\vec{\Psi}$ as
\begin{equation}
||\Delta\vec{\Psi}|| \leq ||(H-E)^-||_{\infty} ||(\Delta E-A)||_{\infty} ||\vec{\Psi}|| \leq 2g^{-1} ||A||_{\infty} ||\vec{\Psi}|| < ||\vec{\Psi}||.
\end{equation}
Then, we determine $\Delta'\vec{\Psi}$ as $\Delta'\vec{\Psi} = \left(\sqrt{1-(||\Delta\vec{\Psi}||^2/||\vec{\Psi}'_0||^2)} - 1\right)\vec{\Psi}'_0+\Delta\vec{\Psi}$, and its norm is bounded by
\begin{eqnarray}
||\Delta'\vec{\Psi}|| &=& \sqrt{\left|\left|\left(1-\sqrt{1-\frac{||\Delta\vec{\Psi}||^2}{||\vec{\Psi}'_0||^2}}\right)\vec{\Psi}'_0\right|\right|^2+||\Delta\vec{\Psi}||^2} \\
&{\leq}& \sqrt{2}||\Delta\vec{\Psi}|| \\
&{\leq}& 2\sqrt{2}g^{-1} ||A||_{\infty} ||\vec{\Psi}||,
\end{eqnarray}
which indicates the existence of the eigenvector $\vec{\Psi}_0$ of $H$ that satisfies $||\vec{\Psi}_0||=||\vec{\Psi}||$ and $||\vec{\Psi}-\vec{\Psi}_0|| \leq D g^{-1}||A||_{\infty}||\vec{\Psi}_0||$ ($D=2\sqrt{2}$).

\subsubsection{Proof of Proposition 1.}
Next, we prove Proposition 1. We assume that the nonlinear eigenvalue problem $f(\vec{\Psi}) = E\vec{\Psi}$ satisfies the following conditions in Theorem: (1) When we rewrite the nonlinear function $f$ as $f(\vec{\Psi})=H(\vec{\Psi})\vec{\Psi}$, there exists a positive real number $c<1$ that satisfies $||H(\vec{\Psi})-H(\vec{0})|| < gc/2$ for any complex vector $\vec{\Psi}$ with $g$ being the bandgap of $H(\vec{0})$ and $||\cdot||$ being the operator norm. (2) There exists a positive real number $c'<1$ such that for any pairs of complex vectors $\vec{\Psi}$ and $\vec{\Psi}+\Delta\vec{\Psi}$ with the norm $w$, they satisfy $||H(\vec{\Psi}+\Delta\vec{\Psi})-H(\vec{\Psi})|| \leq g(1-c)c'(6\sqrt{2}w)^{-1} ||\Delta\vec{\Psi}||$. (3) For any complex vector $\vec{\Psi}$, one can rewrite the nonlinear function $f(\vec{\Psi})$ as $f(\vec{\Psi})=\tilde{H}(\vec{\Psi})\vec{\Psi}$, where $\tilde{H}$ is a Hermitian matrix. (4) The nonlinear equation satisfies the $U(1)$ symmetry, $f(e^{i\theta}\vec{\Psi})=e^{i\theta}f(\vec{\Psi})$. Then, Proposition 1 reads as follows: In the nonlinear eigenvalue problem that satisfies the assumptions in Theorem, the self-consistent calculation converges, $E_i \rightarrow E_{\infty}$, $\vec{\Psi}_i \rightarrow \vec{\Psi}_{\infty}$. Furthermore, there exists an eigenvector $\vec{\Psi}_0$ and eigenvalue $E_0$ of $H(0)$ that satisfy $||E_{\infty}-E_0|| < gc/2(1-c')$ and $||\vec{\Psi}_{\infty}-\vec{\Psi}_0|| < 2^{-1/2}c(1-2c')^{-1}w$.

We denote the eigenvector obtained in the $i$th step of the self-consistent calculation by $\vec{\Psi}_i$. Mathematically, $\vec{\Psi}_i$ is defined as follows: $\vec{\Psi}_0$ is an eigenvector of $H(\vec{0})$, $H(\vec{0})\vec{\Psi}_0 = E_0\vec{\Psi}_0$. $\vec{\Psi}_i$ ($i=1,2,\cdots$) is an eigenvector of $H(\vec{\Psi}_{i-1})$ whose eigenvalue is the closest to the eigenvalue of $\vec{\Psi}_{i-1}$ and that minimizes the distance between $\vec{\Psi}_{i-1}$ and $\vec{\Psi}_{i}$. We also denote the band gap of $H(\vec{\Psi}_i)$ as $g_{i+1}$. 

To prove Proposition 1, we iteratively use the perturbation theorem of linear eigenvectors and evaluate the distance between $\vec{\Psi}_{i-1}$ and $\vec{\Psi}_{i}$. In the following, we explain how to evaluate the distance at each step. From the first condition, we obtain an inequality $||H(\vec{\Psi}_0)-H(\vec{0}))||<gc/2$. Then combining the inequality and the perturbation theorem, we show the upper bound of the distance between $\vec{\Psi}_{0}$ and $\vec{\Psi}_{1}$, $||\vec{\Psi}_1-\vec{\Psi}_0||<\sqrt{2}cw$. The linear perturbation theorem \cite{Bhatia1997} also indicates $g_1 > g(1-c)$. We next use the second condition, which indicates $||H(\vec{\Psi}_1)-H(\vec{\Psi}_0))||<(1-c)cc'g/6$. Combining these inequalities and the perturbation theorem of linear eigenvectors, we obtain 
\begin{eqnarray}
||\vec{\Psi}_2-\vec{\Psi}_1|| &<& 2\sqrt{2}w(1-c)cc'g/(6g_1) \nonumber\\
&<& \frac{\sqrt{2}}{3}(1-c)cc'w \nonumber\\
&<& \frac{\sqrt{2}}{3}cc'w, \label{bound1}
\end{eqnarray}
 where we use $0<c<1$ in the last inequality. From the conditions $c<1$ and $c'<(1-c)/2<1$ and the linear perturbation theorem, we also obtain $g_2 > g(1-c)(1-cc'/3) > 2g(1-c)/3$. We again use the second condition and derive $||H(\vec{\Psi}_2)-H(\vec{\Psi}_1))||<(1-c)c(c'/2)^2g/6$. From the perturbation theorem, we obtain 
\begin{eqnarray}
 ||\vec{\Psi}_3-\vec{\Psi}_2|| &<& 2\sqrt{2}(1-c)c\left(\frac{c'}{2}\right)^2\frac{g}{6g_2} \nonumber\\ 
 &<& \frac{4\sqrt{2}cw}{9}\left(\frac{c'}{2}\right)^2 \nonumber\\
 &<& \frac{cw}{\sqrt{2}}\left(\frac{c'}{2}\right)^2. \label{bound2}
\end{eqnarray}
Then, we obtain the bounds $g_3 > g(1-c)[1-cc'/3-c(c'/2)^2/3]$ and $||H(\vec{\Psi}_3)-H(\vec{\Psi}_2))||<(1-c)c(c'/2)^3g/6$. To continue the calculations, we utilize the inequality $g(1-c)[1-cc'/3-(c/3)\sum_{n=2}^i (c'/2)^n]> g(1-c)/2$. By iterating the evaluation of the upper bound of the band gap $g_i$ and the distance between $\vec{\Psi}_{i-1}$ and $\vec{\Psi}_{i}$, we obtain 
\begin{eqnarray}
&{}& g_i > g(1-c)\left[1-\frac{cc'}{3}-\frac{c(c')^2}{12(1-c'/2)}\right], \label{bound_eval}\\
 &{}& ||\vec{\Psi}_i-\vec{\Psi}_{i-1}|| < \frac{(1-c)cw}{\sqrt{2}}\left(\frac{c'}{2}\right)^{i-1}.
\end{eqnarray}

The above inequality indicates that the distance between $\vec{\Psi}_i$ and $\vec{\Psi}_0$ is also bounded by a constant
\begin{eqnarray}
 ||\vec{\Psi}_n-\vec{\Psi}_0|| &{\leq}& \sum_{i=0}^{n-1} ||\vec{\Psi}_i-\vec{\Psi}_{i+1}|| \nonumber\\
  &<& cw \left[ \sqrt{2} +\frac{\sqrt{2}(1-c)c'}{3} + \frac{(1-c)(c')^2}{4\sqrt{2}(1-c'/2)} \right]. \label{bound_evec}
\end{eqnarray}
Therefore, this infinite series is a Cauchy sequence, and thus $\vec{\Psi}_i$ converges to $\lim_{i\rightarrow\infty} \vec{\Psi}_i$. Supplementary Equation \eqref{bound_evec} also presents the upper bound of the distance between $\lim_{i\rightarrow\infty} \vec{\Psi}_i$ and $\vec{\Psi}_0$. We can also obtain the bound of the difference between the nonlinear eigenvalue and the corresponding linear eigenvalue from half of the right-hand side of Supplementary Eq.~\eqref{bound_eval}.

We note that we have loosely (but rigorously) evaluated the bound in Supplementary Eqs.~\eqref{bound1} and \eqref{bound2}. To obtain the tight inequality, we need to solve a complicated recurrence relation and thus we leave it to future works. While the obtained bound is loose, Proposition 1 still guarantees that the algorithm of the self-consistent calculation can solve the nonlinear eigenvalue problem in weakly nonlinear systems. It is also noteworthy that one can also prove the convergence of the self-consistent calculation in the case $(1-c)/2 \leq c'<1$ in the completely same way. The condition of $c'<(1-c)/2$ is needed only for proving Lemma.

\subsubsection{Proof of Proposition 2.}
While Proposition 1 guarantees the convergence of the self-consistent calculation, there still remains a possibility that the convergent solution is not a pair of genuine nonlinear eigenvalue and eigenvector. Thus, we need to prove Proposition 2, which guarantees that the obtained solution satisfies the nonlinear eigenequation. 

{\it Proposition 2:} The converged solutions of the self-consistent calculation, $E_{\infty}$ and $\vec{\Psi}_{\infty}$ satisfy $f(\vec{\Psi}_{\infty}) = E_{\infty}\vec{\Psi}_{\infty}$.

Let $\vec{\Psi}_i$ and $E_i$ be the obtained solution of an eigenvalue and an eigenvector at the $i$th step of the self-consistent calculation, respectively. We also consider their limits, $\vec{\Psi} = \lim_{i\rightarrow\infty} \vec{\Psi}_i$ and $E = \lim_{i\rightarrow\infty} E_i$. Then, Proposition 2 is equivalent to $H(\vec{\Psi})\vec{\Psi} - E \vec{\Psi} = 0$. The norm of the left-hand side is upper-bounded by
\begin{eqnarray}
&{}& ||H(\vec{\Psi}) \vec{\Psi} - E \vec{\Psi}|| \nonumber\\
&{\leq}& ||H(\vec{\Psi}) (\vec{\Psi} - \vec{\Psi}_n)|| + ||[H(\vec{\Psi}) - H(\vec{\Psi}_{n-1})] \vec{\Psi}_n|| + ||E_n \vec{\Psi}_n - E \vec{\Psi}|| + ||H(\vec{\Psi}_{n-1}) \vec{\Psi}_n - E_n \vec{\Psi}_n|| \nonumber\\
&=& ||H(\vec{\Psi}) (\vec{\Psi} - \vec{\Psi}_n)|| + ||[H(\vec{\Psi}) - H(\vec{\Psi}_{n-1})] \vec{\Psi}_n|| + ||E_n \vec{\Psi}_n - E \vec{\Psi}||. \label{bound_eigeneq}
\end{eqnarray}
Since we assume that $H(\vec{\Psi})$ is always a bounded operator, the first term converges to zero. By using the second condition in Theorem, we obtain the upper bound of the second part as $||[H(\vec{\Psi}) - H(\vec{\Psi}_{n-1})] \vec{\Psi}_n|| < (1-c)cc'g_n ||\vec{\Psi}-\vec{\Psi}_{n-1}||\, ||\vec{\Psi}_n||/(6\sqrt{2})$ with $g_n$ being the minimum difference of the nonlinear eigenvalues of  $H(\vec{\Psi}_{n-1})$. Because of $\lim_{n\rightarrow\infty} ||\vec{\Psi}-\vec{\Psi}_{n-1}||=0$, the second term also converges to zero. Lastly, one can also check the convergence of the third term to zero from the definitions of $E$ and $\vec{\Psi}$. Therefore, the right-hand side of Supplementary Eq.~\eqref{bound_eigeneq} converges to zero, which deduces that the left-hand side also converges to zero in the limit of $n\rightarrow\infty$.

\subsubsection{Proof of Lemma.}
The proof of Lemma is similar to that of Proposition 1. The major difference is that we use the nonlinear eigenvector $\vec{\Psi}$ of $H+f(\vec{\Psi})$ as the initial guess of the nonlinear eigenvector of $H+(1-\epsilon)f(\vec{\Psi})$. In this subsection, we define $\vec{\Psi}'_i$ as follows: The initial guess $\vec{\Psi}'_0$ is $\vec{\Psi}'_0 = \vec{\Psi}$. The obtained solution at the $i$th step $\vec{\Psi}'_i$ is the eigenvector of $H+(1-\epsilon)f(\vec{\Psi}'_{i-1})$, $[H+(1-\epsilon)f(\vec{\Psi}'_{i-1})]\vec{\Psi}'_{i} = E_i \vec{\Psi}'_{i}$, whose eigenvalue $E'_i$ is the closest to that of the previous solution $E'_{i-1}$. We also describe the minimum difference of the eigenvalues of $H+f(\vec{\Psi})$ by $g'$, which is lower-bounded by Supplementary Eq.~\eqref{bound_eval}. The minimum difference of the eigenvalues of $H+(1-\epsilon)f(\vec{\Psi}'_{i-1})$ is denoted by $g'_i$

Comparing $H+f(\vec{\Psi})$ and $H+(1-\epsilon)f(\vec{\Psi})$, we obtain the following inequality,
\begin{eqnarray}
||[H+f(\vec{\Psi})]-[H+(1-\epsilon)f(\vec{\Psi})]||_{\infty} &=& \epsilon||H(\vec{\Psi})-H(\vec{0})||_{\infty} \nonumber\\
&<& \frac{cg\epsilon}{2} < \frac{cg'}{2}.
\end{eqnarray}
Then, we obtain the bound of $g'_1$ and $||\vec{\Psi}'_1-\vec{\Psi}'_0||$,
\begin{eqnarray}
&{}& g_1 > g'\left( 1- \frac{cg\epsilon}{g'}\right) > g'(1-c), \\
&{}& ||\vec{\Psi}'_1-\vec{\Psi}'_0|| < \sqrt{2} \frac{g}{g'} c\epsilon w.
\end{eqnarray}
As in the proof of Proposition 1, we iteratively evaluate the bound of $||H(\vec{\Psi}'_i)-H(\vec{\Psi}'_{i-1})||$, $g'_i$, and $||\vec{\Psi}'_i-\vec{\Psi}'_{i-1}||$. We finally obtain
\begin{eqnarray}
&{}& g'_i > g'(1-c)\left[1-\frac{gcc'\epsilon}{3g'}-\sum_{j=2}^{i-1} \frac{c\epsilon}{3} \left(\frac{gc'}{2g'}\right)^j \right], \label{bound_eval2}\\
 &{}& ||H(\vec{\Psi}'_i)-H(\vec{\Psi}'_{i-1})||_{\infty} < \frac{(1-c)cg\epsilon}{6}\left(\frac{gc'}{2g'}\right)^{i}, \\
 &{}& ||\vec{\Psi}'_i-\vec{\Psi}'_{i-1}|| < \frac{(1-c)cg\epsilon w}{\sqrt{2}g'}\left(\frac{gc'}{2g'}\right)^{i-1}, \label{bound_evec2}
\end{eqnarray}
for $i\geq 3$. From the condition of $c'<(1-c)/2$, we obtain $gc'/2g'<1$. Therefore, Supplementary Equation \eqref{bound_evec2} indicates the convergence of the self-consistent calculation. We can also derive the upper bound of the distance between $\vec{\Psi}$ and $\vec{\Psi}_{\epsilon}$,
\begin{eqnarray}
||\vec{\Psi} - \vec{\Psi}_{\epsilon}|| < cw\epsilon \left[ \frac{\sqrt{2}g}{g'} +\frac{\sqrt{2}(1-c)g^2c'}{3(g')^2} + \frac{(1-c)g^3(c')^2}{4\sqrt{2}(g')^3(1-\frac{gc'}{2g'})} \right],
\end{eqnarray}
whose right-hand side is the product of $\epsilon$ and a constant that is independent of the eigenvector $\vec{\Psi}$.

\subsubsection{Proof of Theorem.}
To prove the bulk-boundary correspondence in weakly nonlinear systems that satisfies the conditions in Theorem, we show that the nonlinear eigenvectors can be continuously connected to linear eigenvectors. Here, we consider the nonlinear eigenvectors $\vec{\Psi}_{\epsilon}$ of $H+\epsilon f(\vec{\Psi})$ ($0\leq\epsilon\leq1$) whose nonlinear eigenvalues are the closest to a certain eigenvalue $E$ of $H$ of all the nonlinear eigenvalues. Since we assume that the number of nonlinear bands is equal to that of linear bands of $H$, the nonlinear eigenvector of $H+\epsilon f(\vec{\Psi})$ should match that obtained from the self-consistent calculation starting from the nonlinear eigenvector of $H+\epsilon f(\vec{\Psi}) / (1 - \delta)$, which is considered in Lemma. Therefore, Lemma indicates that $\vec{\Psi}_{\epsilon}$ is a continuous vector function of $\epsilon$ and thus provides a continuous path connecting the nonlinear eigenvector of $H+ f(\vec{\Psi})$ and the linear eigenvector of $H$. 

Since the nonlinear Chern number is a topological invariant that cannot change its value under the continuous deformation of the nonlinear eigenvectors, the nonlinear Chern number coincides with the linear Chern number of $H$. One can also construct adiabatic connections between the edge modes of linear systems and nonlinear eigenvectors in nonlinear systems with open boundaries and nonzero nonlinear Chern numbers. Therefore, the bulk-boundary correspondence in weakly nonlinear systems is proved from that in linear systems.

\subsection{Discretization of the nonlinear Dirac Hamiltonian.} To confirm the bulk-boundary correspondence in the nonlinear QWZ model with stronger nonlinearity, we discretize the nonlinear Dirac Hamiltonian in Eq.~\eqref{nonlinearDirac}. Naively taking the lattice constant $h$, one seems to achieve the discretization of the Dirac Hamiltonian,
\begin{eqnarray}
H(\Psi) = \left(
  \begin{array}{cc}
   m + \kappa (|\Psi_1|^2+|\Psi_2|^2) & (i\Delta_x - \Delta_y)/h \\
   (i\Delta_x + \Delta_y)/h & -(m + \kappa (|\Psi_1|^2+|\Psi_2|^2))
  \end{array}
  \right),
\end{eqnarray}
where $\Delta_i$ are the difference operators defined as $\Delta_i \Psi(\mathbf{x})=[\Psi(\mathbf{x}+h\mathbf{e}_i)-\Psi(\mathbf{x}-h\mathbf{e}_i)]/2$. However, at the gapless parameter $m + \kappa (|\Psi_1|^2+|\Psi_2|^2)=0$, the obtained lattice Hamiltonian closes gaps at four points in the wavenumber space $(k_x,k_y)=(0,0),(0,\pi),(\pi,0),(\pi,\pi)$, which have different signs of topological charges around them. Thus, we cannot obtain the expected topological features such as the nonzero Chern number and the localized modes in this naive discretization. To avoid this inconsistency, we introduce the wavenumber-dependent mass terms $\Delta^2/h$, where $\Delta^2$ is the discretized Laplacian, $\Delta^2 \Psi(x,y)=\Psi(x+h,y)+\Psi(x-h,y)+\Psi(x,y+h)+\Psi(x,y-h)-4\Psi(x,y)$. The added mass terms correspond to the Wilson fermion action \cite{Wilson1974,Wilson1977} in the lattice field theory. We finally obtain the discretized Hamiltonian, which is denoted as 
\begin{eqnarray}
H(\Psi) = \left(
  \begin{array}{cc}
   m + \kappa w + (\cos k_x+\cos k_y)/h  & (\sin k_x + i\sin k_y)/h \\
   (\sin k_x - i\sin k_y)/h & -[m + \kappa w +(\cos k_x+\cos k_y)/h] 
  \end{array}
  \right),
  \label{discretized-Dirac}
\end{eqnarray}
in the wavenumber space, where $w$ is the squared amplitude $w=|\Psi_1|^2+|\Psi_2|^2$. As expected from the derivation of the nonlinear Dirac Hamiltonian in the main text, this discretized Hamiltonian is equivalent to that of the nonlinear QWZ model in Eq.~\eqref{model-wavenumber} except for the existence of the lattice constant $h$. In the $h\rightarrow 0$ limit, the obtained lattice Hamiltonian reproduces the behavior of the nonlinear Dirac Hamiltonian.

\subsection{Numerical calculations of the quench dynamics.}
We have numerically solved the nonlinear Schr\"{o}dinger equation (Eq.~\eqref{model-real}) with the initial condition localized at the left edge as shown in Fig.6a. As the initial conditions, we study the two cases. One is
\begin{equation}
\Psi _{1}\left( x,y\right) =\delta _{x,1},\qquad \Psi _{2}\left( x,y\right)
=i\delta _{x,1}, \label{initial_condition1}
\end{equation}
which corresponds to the solution with $u=-1$ of the linear model ($\kappa =0$). The other is 
\begin{equation}
\Psi _{1}\left( x,y\right) =\left( -1\right) ^{y}\delta _{x,1},\qquad \Psi_{2}\left( x,y\right) =i\left( -1\right) ^{y}\delta _{x,1}, \label{initial_condition2}
\end{equation}
which corresponds to the solution with $u=1$ of the linear model ($\kappa =0$). The derivation is shown in Supplementary Note 10. For the numerical calculation, we used the ``NDSolve'' function in Mathematica. We consider a $10\times10$ lattice under the open boundary condition in the $x$ direction and the periodic boundary condition in the $y$ direction.

We solve the nonlinear Schr\"{o}dinger equation (Eq.~\eqref{model-real}) under the two initial conditions in Eqs.~\eqref{initial_condition1} and \eqref{initial_condition2}. The time evolution of $\left\vert \Psi_{1}\left( x,y\right) \right\vert ^{2}+\left\vert \Psi _{2}\left( x,y\right)
\right\vert ^{2}$ is shown in Fig.6d and e. We define the phase indicator $P$ by
\begin{equation}
P\equiv \max_{0.99T<t<T}\left[ \left\vert \Psi _{1}\left( x,y\right)
\right\vert ^{2}+\left\vert \Psi _{2}\left( x,y\right) \right\vert ^{2}\right] 
\end{equation}
with $T=10$. We have $P\sim0$ in the trivial phase as shown in Fig.6b, which implies the absence of localized edge states. On the other hand, we have $P\sim1$ in a topological phase as shown in Fig.6c, which implies the presence of a localized edge mode. In order to elucidate the topological phase diagram, we plot $P$ in the $u$-$\kappa w$ plane in Fig.6d and e. We find that the phase indicator is 1 along the blue line, where the exact solution is valid. It shows that the exact solution is realized by the quench dynamics.

\subsection{Derivation of a nonlinear eigenvalue problem in the wavenumber space.}
We use the wavenumber-space description of a nonlinear eigenequation to introduce the nonlinear Chern number. Here, we clarify the relationship between the real-space and wavenumber-space description of the nonlinear eigenequation. For clarity, we consider a one-dimensional lattice system with translation symmetry and a periodic boundary. In general, one can describe the eigenequation of such a one-dimensional nonlinear system as
\begin{equation}
 E\vec{\psi}(x) = \vec{F}(\vec{\psi}(x),\vec{\psi}(x+1),\cdots,\vec{\psi}(L),\vec{\psi}(1),\cdots,\vec{\psi}(x-1)),
 \label{eigen_eq}
\end{equation}
where $E$ and $\vec{\Psi} = (\vec{\psi}(1),\cdots,\vec{\psi}(x-1),\vec{\psi}(x),\vec{\psi}(x+1),\cdots,\vec{\psi}(L))^T$ with $\vec{\psi}(x)$ being $\vec{\psi}(x)=(\psi_1(x),\psi_2(x),\cdots,\psi_n(x))$ represent the nonlinear eigenvalue and eigenvector, respectively. Due to the periodicity, $\vec{F}$ does not explicitly depend on the location $x$.

To derive the wavenumber-space description of the nonlinear eigenequation, we have assumed the Bloch ansatz,
\begin{equation}
\vec{\psi}(x) = e^{ikx} \vec{\psi}',
\end{equation}
with $k$ being $k=2\pi n/L$, $n\in\{0,1,2,\cdots,L-1\}$. We have also assumed the $U(1)$ symmetry denoted by 
\begin{equation}
\vec{F}(e^{i\theta}\vec{\psi}(x),\cdots,e^{i\theta}\vec{\psi}(x-1)) = e^{i\theta}\vec{F}(\vec{\psi}(x),\cdots,\vec{\psi}(x-1)),
\end{equation}
where $\theta$ is a real phase factor. Substituting the Bloch wave ansatz in Supplementary Eq.~\eqref{eigen_eq}, we obtain
\begin{eqnarray}
 Ee^{ikx}\vec{\psi}' &=& \vec{F}(e^{ikx}\vec{\psi}',e^{ik(x+1)}\vec{\psi}',\cdots,e^{ikL}\vec{\psi}',e^{ik}\vec{\psi}',\cdots,e^{ik(x-1)}\vec{\psi}') \nonumber\\
 &=& e^{ikx} \vec{F}(\vec{\psi}',e^{ik}\vec{\psi}',\cdots,e^{ik(L-1)}\vec{\psi}').
 \label{eigen_eq2}
\end{eqnarray}
Then, we finally derive the wavenumber-space description of the nonlinear eigenequation
\begin{eqnarray}
 E\vec{\psi}' = \vec{F}(\vec{\psi}',e^{ik}\vec{\psi}',\cdots,e^{ik(L-1)}\vec{\psi}') =: \vec{F}(\vec{\psi}',k),
 \label{eigen_eq3}
\end{eqnarray}
where both sides are independent of the location $x$.

One can also derive the wavenumber-space description of the nonlinear eigenequation corresponding to a continuum nonlinear system (cf. the nonlinear Dirac Hamiltonian in Eq.~(7) in the main text). When a nonlinear system has continuous translation symmetry, the real-space description of an eigenequation is 
\begin{equation}
 E\vec{\psi}(x) = \vec{F}(\vec{\psi}(x),\partial_x \vec{\psi}(x),\partial_x^2\vec{\psi}(x),\cdots)
 \label{eigen_eq_cont}
\end{equation}
in a general continuum nonlinear system. To derive its wavenumber-space description, one should assume the Bloch-wave ansatz,
\begin{equation}
\vec{\psi}(x) = e^{ikx} \vec{\psi}'.
\end{equation}
We also impose the $U(1)$ symmetry described as 
\begin{equation}
\vec{F}(e^{i\theta}\vec{\psi}(x),e^{i\theta}\partial_x \vec{\psi}(x),e^{i\theta}\partial_x^2\vec{\psi}(x),\cdots) = e^{i\theta}\vec{F}(\vec{\psi}(x),\partial_x \vec{\psi}(x),\partial_x^2\vec{\psi}(x),\cdots) .
\end{equation}
Then, the eigenequation (Supplementary Eq.~\eqref{eigen_eq_cont}) reads 
\begin{eqnarray}
 Ee^{ikx}\vec{\psi}' &=& \vec{F}(e^{ikx}\vec{\psi}',\partial_x [e^{ikx}\vec{\psi}'],\partial_x^2[e^{ikx}\vec{\psi}'],\cdots) \nonumber\\
 &=& \vec{F}(e^{ikx}\vec{\psi}',ike^{ikx} \vec{\psi}',-k^2e^{ikx} \vec{\psi}',\cdots) \nonumber\\
 &=& e^{ikx} \vec{F}(\vec{\psi}',ik \vec{\psi}',-k^2 \vec{\psi}',\cdots).
 \label{eigen_eq_cont2}
\end{eqnarray}
Therefore, one can obtain the wavenumber-space description of the eigenequation
\begin{eqnarray}
 E\vec{\psi}' =  \vec{F}(\vec{\psi}',ik \vec{\psi}',-k^2 \vec{\psi}',\cdots) =: \vec{F}(\vec{\psi}',k).
 \label{eigen_eq_cont3}
\end{eqnarray}

In continuum systems with discrete translation symmetry, the real-space description of an eigenequation is 
\begin{equation}
 E\vec{\psi}(x) = \vec{F}(x;\vec{\psi}(x),\partial_x \vec{\psi}(x),\partial_x^2\vec{\psi}(x),\cdots).
 \label{eigen_eq_cont_a}
\end{equation}
We assume the Bloch ansatz
\begin{equation}
\vec{\psi}(x) = e^{ikx} \vec{\psi}'(x),
\end{equation}
where $\vec{\psi}'(x)$ is a periodic function with the same period as that of $\vec{F}(x;\cdot)$. The $U(1)$ symmetry is described as 
\begin{equation}
\vec{F}(x;e^{i\theta}\vec{\psi}(x),e^{i\theta}\partial_x \vec{\psi}(x),e^{i\theta}\partial_x^2\vec{\psi}(x),\cdots) = e^{i\theta}\vec{F}(x;\vec{\psi}(x),\partial_x \vec{\psi}(x),\partial_x^2\vec{\psi}(x),\cdots) 
\end{equation}
Then, we rewrite the eigenequation (Supplementary Eq.~\eqref{eigen_eq_cont_a}) as
\begin{eqnarray}
 Ee^{ikx}\vec{\psi}'(x) &=& \vec{F}(x;e^{ikx}\vec{\psi}'(x),\partial_x [e^{ikx}\vec{\psi}'(x)],\partial_x^2[e^{ikx}\vec{\psi}'(x)],\cdots) \\
 &=& \vec{F}(x;e^{ikx}\vec{\psi}'(x),e^{ikx}(\partial_x+ik) \vec{\psi}'(x),e^{ikx}(\partial_x+ik)^2 \vec{\psi}'(x),\cdots) \\
 &=& e^{ikx} \vec{F}(x;\vec{\psi}'(x),(\partial_x+ik)  \vec{\psi}'(x),(\partial_x+ik)^2 \vec{\psi}'(x),\cdots).
 \label{eigen_eq_cont_a2}
\end{eqnarray}
Therefore, we obtain the wavenumber-space description of the eigenequation
\begin{eqnarray}
 E\vec{\psi}'(x) = \vec{F}(x;\vec{\psi}'(x),(\partial_x+ik)  \vec{\psi}'(x),(\partial_x+ik)^2 \vec{\psi}'(x),\cdots) =: \vec{F}(x;\vec{\psi}',k).
 \label{eigen_eq_cont_a3}
\end{eqnarray}
In both lattice and continuum cases, the size of the eigenvector in the wavenumber space is smaller than that in the real space, which is of practical advantage in the calculation of the nonlinear Chern number. We also note that one can derive the wavenumber-space description of nonlinear eigenequations in higher dimensions via the same calculations as one-dimensional cases.

In linear systems, one can derive the wavenumber-space descriptions of Hamiltonians from the Fourier transformation of the linear equations. In contrast, the Fourier transformation of a nonlinear equation reveals the interaction between eigenmodes with different wavenumbers. For example, in the Fourier transformation of a Kerr-type nonlinearity, $|\psi(x)|^2\psi(x)$ becomes
\begin{eqnarray}
 \mathcal{F}(|\psi(x)|^2\psi(x)) (k) &=& \frac{1}{2\pi} \int dx e^{-ikx} |\psi(x)|^2\psi(x) \nonumber\\
 &=& \frac{1}{2\pi} \int dx e^{-ikx} \int dk_1 dk_2 dk_3 e^{-ik_1x} \psi^{\ast}(k_1) e^{ik_2x} \psi(k_2) e^{ik_3x} \psi(k_3) \nonumber\\
 &=& \frac{1}{2\pi} \int dx \int dk_1 dk_2 dk_3 e^{i(-k-k_1+k_2+k_3)x}  \psi^{\ast}(k_1) \psi(k_2) \psi(k_3) \nonumber\\
 &=& \frac{1}{2\pi} \int dk_1 dk_2 dk_3 \delta(-k-k_1+k_2+k_3)  \psi^{\ast}(k_1) \psi(k_2) \psi(k_3) \nonumber\\
 &=& \frac{1}{2\pi} \int dk_1 dk_2 \psi^{\ast}(k_1) \psi(k_2) \psi(k+k_1-k_2). \label{Fourier-Kerr}
\end{eqnarray}
However, since we assume the Bloch-wave ansatz $\vec{\psi}(x) = e^{ikx} \vec{\psi}'$ or equivalently the twisted boundary condition in a unit cell in the derivation of the wavenumber-space description of the nonlinear eigenequation, the integrant in Supplementary Eq.~\eqref{Fourier-Kerr} contributes only when the wavenumbers are $k_1=k_2 = k$ such that $\mathcal{F}(|\psi(x)|^2\psi(x)) (k)=|\psi(k)|^2\psi(k)$.

\subsection{Quantization of the nonlinear Chern number.}
The nonlinear Chern number defined in Eq.~(3) in the main text are quantized. We can show the quantization by embedding the nonlinear eigenvectors into eigenspaces of a linear Bloch Hamiltonian. For clarity, we consider two-band cases in the following. 

\begin{figure*}
  \includegraphics[width=140mm,bb=0 0 1245 615,clip]{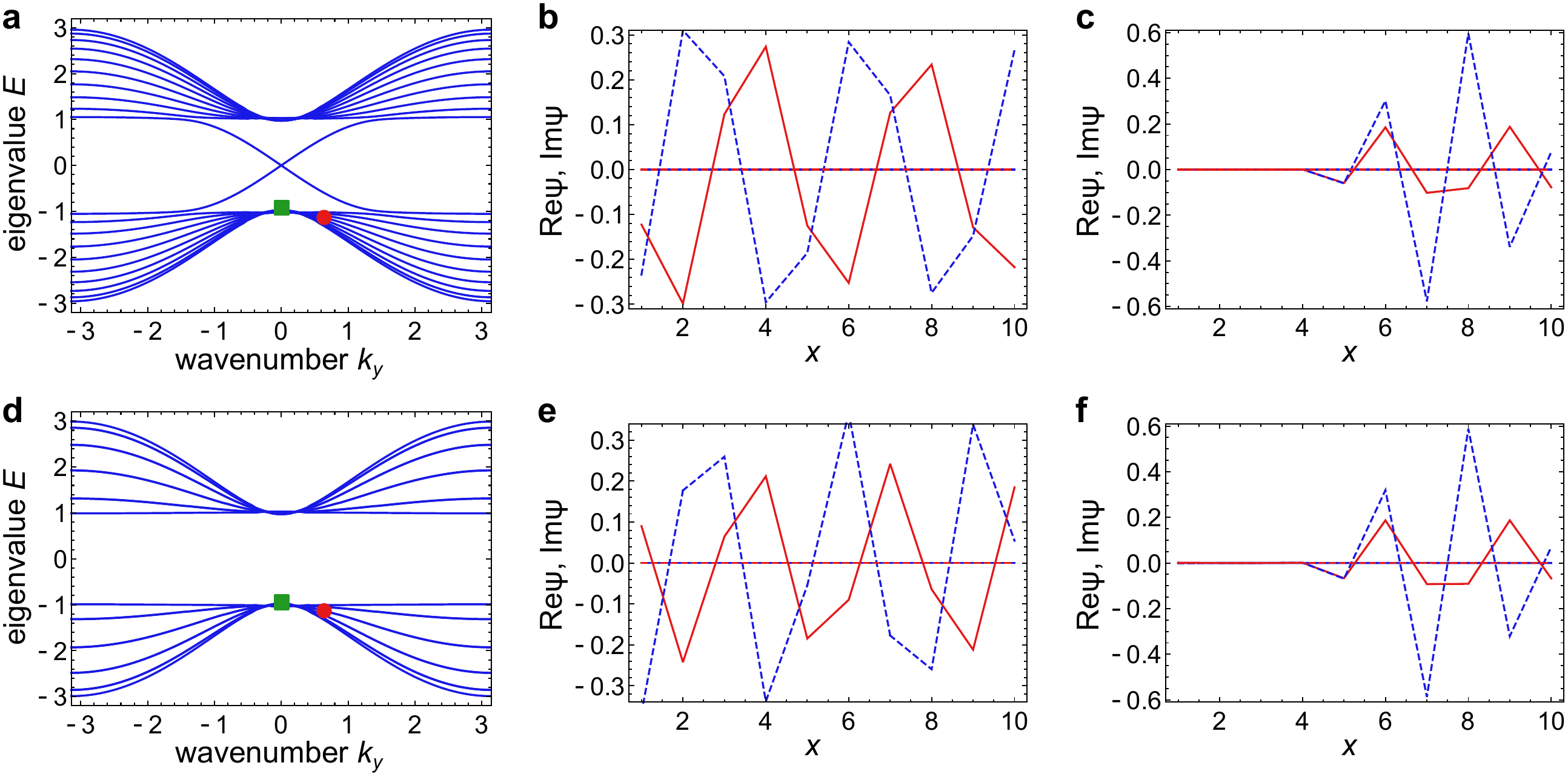}
  \caption{\label{supplefig1} {\bf Bulk nonlinear eigenvectors of the nonlinear Qi-Wu-Zhang (QWZ) model.} {\bf a,} Nonlinear band structures under the open boundary condition in the $x$ direction and the twisted boundary condition in the $y$ direction. The red circle and the green square correspond to the nonlinear eigenvectors in panels {\bf b} and {\bf c}, respectively. The parameters used are $u=-0.1$ and $\kappa w=0.1$. {\bf b,} Wave function of a sine-wave bulk mode under the open boundary condition in the $x$ direction. The red (blue) lines show the real part of the first (second) component of the wave function, and the red (blue) dashed lines show the imaginary part of the first (second) component (these correspondences are the same in panels {\bf c, e,} and {\bf f}). {\bf c,} Wave function of a localized bulk mode under the open boundary condition in the $x$ direction. The localization indicates that there can be eigenstates out of the Bloch-wave ansatz. {\bf d,} Nonlinear band structures under the periodic boundary condition in the $x$ direction. We consider a $10\times1$ supercell structure and impose the twisted boundary condition in the $y$ direction. The red circle and the green square correspond to the nonlinear eigenvectors in panels {\bf e} and {\bf f}, respectively.  {\bf e,} Wave function of a sine-wave bulk mode under the periodic boundary condition in the $x$ direction. {\bf f,} Wave function of a localized bulk mode under the open boundary condition in the $x$ direction. The localization and periodicity are not altered by the boundary condition.}
\end{figure*}

Let $\psi(\mathbf{k})$ be the nonlinear eigenvectors of the nonlinear eigenvalue equation $F(\psi)=E\psi$ derived from the Bloch ansatz of the wavenumber $\mathbf{k}$. Then, one can uniquely determine $\phi(\mathbf{k})$ so that $\phi(\mathbf{k})$ is perpendicular to $\psi(\mathbf{k})$ (we here ignore the phase ambiguity). By using $\phi(\mathbf{k})$ and $\phi(\mathbf{k})$, one can construct the following linear Bloch Hamiltonian
\begin{equation}
H(\mathbf{k}) = -|\psi(\mathbf{k})\rangle \langle \psi(\mathbf{k})| + |\phi(\mathbf{k})\rangle \langle \phi(\mathbf{k})|.
\end{equation}
This Bloch Hamiltonian has the eigenvector $\psi(\mathbf{k})$ and the corresponding eigenvalue $E(\mathbf{k})=-1$. Therefore, the Chern number of the lower band of this linear Hamiltonian is
\begin{equation}
C = \frac{1}{2\pi i} \int \nabla_{\mathbf{k}} \times \langle \psi(\mathbf{k})| \nabla_{\mathbf{k}} |\psi(\mathbf{k}) \rangle d^2\mathbf{k},
\end{equation}
which is equivalent to the nonlinear Chern number of $F(\psi)=E\psi$. Therefore, the nonlinear Chern numbers must be integers as linear ones. We note that in many-band cases, one can also construct a linear Bloch Hamiltonian that has the same eigenvectors, and thus the nonlinear Chern number must be quantized.

\subsection{Amplitude distributions of nonlinear bulk modes.}
Here, we numerically calculate the nonlinear eigenstates of the nonlinear Qi-Wu-Zhang (QWZ) model (Eq.~(4) in the main text) of nonlinear Chern insulators as in Fig.~3 in the main text. Supplementary Figure \ref{supplefig1} shows an amplitude distribution of a nonlinear bulk eigenstate. The bulk eigenstate at $k_y=\pi/5$ resembles sine curve, which can justify the use of the Bloch-wave ansatz to approximate the bulk modes and derive the nonlinear Chern number in this parameter region. In contrast, the bulk modes around $k_y=0$ are localized as shown in Supplementary Figs.~\ref{supplefig1}c,f because of the nonlinear interactions between linear bulk modes with different wavenumbers. These results are consistent with Proposition 1, which indicates that the nonlinear eigenstates are almost unchanged from the corresponding linear modes in weakly nonlinear regimes, while stronger nonlinearity than the linear band gap can lead to drastic changes in nonlinear eigenvectors. 

While numerical techniques cannot stably achieve periodic solutions around $k_y=0$, periodic solutions still exist in such a stronger nonlinearity region. To capture the topological properties of nonlinear systems, we focus only on such periodic solutions. We also note that weak nonlinearity has a negligible impact on the existence of gapless edge modes because their eigenvalues are isolated from those of bulk modes.

\subsection{Numerical calculation of the Chern number in the nonlinear QWZ model.}
In this section, we numerically demonstrate the quantization and the amplitude dependence of the nonlinear Chern number. We calculate the nonlinear eigenvectors of the wavenumber-space description of the nonlinear QWZ model (Eq. (4) in the main text) as in Fig.~3 in the main text. Then, applying the Fukui-Hatsugai-Suzuki method \cite{Fukui2005} to the obtained nonlinear eigenvectors, we calculate the nonlinear Chern number. 

\begin{figure*}
  \includegraphics[width=140mm,bb=0 0 890 300,clip]{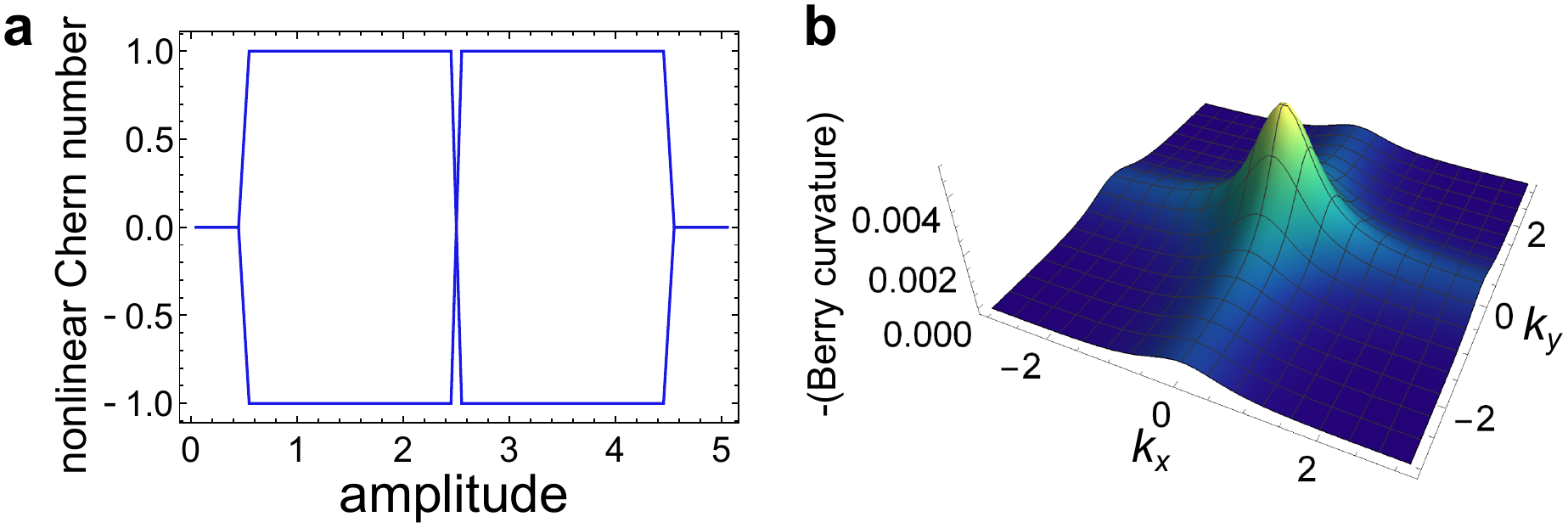}
  \caption{\label{supplefig2} {\bf Nonlinear Chern number and Berry curvature of the nonlinear QWZ model.} {\bf a,} We numerically calculate the nonlinear Chern number of the nonlinear QWZ model at different amplitudes. We plot two nonlinear Chern numbers obtained from the upper and lower nonlinear bands for each amplitude. We can confirm the change of the nonlinear Chern number by the amplitude, which indicates the nonlinearity-induced topological phase transitions. We use the parameters $u=-2.5$, $\kappa=1$. {\bf b,} We plot the nonlinear Berry curvature of the lower nonlinear band of the nonlinear QWZ model at $u=-2.5$ and $\kappa w=1.0$. For clarity, we inverse the sign of the nonlinear Berry curvature. We can confirm the localization of the nonlinear Berry curvature around $(k_x,k_y) = (0,0)$, where the nonlinear band closes a gap at the critical amplitude.}
\end{figure*}

Supplementary Figure \ref{supplefig2}a shows the result of the numerical calculation of the nonlinear Chern number. We confirm that the nonlinear Chern numbers always take integer values as in linear cases. The figure also indicates the existence of the nonlinearity-induced topological phase transitions where the nonlinear Chern numbers are altered at the critical amplitudes $w=0.5,\,2.5,\,4.5$. Supplementary Figure \ref{supplefig2}b presents the nonlinear Berry curvature 
\begin{equation}
\nabla_{\mathbf{k}} \times \left(\frac{1}{\sqrt{w}}\langle \Psi_j(\mathbf{k})|\right) \nabla_{\mathbf{k}} \left(\frac{1}{\sqrt{w}}|\Psi_j(\mathbf{k}) \rangle\right),
\end{equation}
whose integration is equal to the nonlinear Chern number. The nonlinear Berry curvature is localized at $\mathbf{k}=(0,0)$ where the nonlinear bands become gapless at the critical parameters.

\subsection{Nonlinear Dirac Hamiltonians derived from low-energy expansions of the nonlinear QWZ model.}
We derive the nonlinear Dirac Hamiltonian in Eq.~(7) in the main text from the small-wavenumber expansion of the nonlinear QWZ model around $(k_x,k_y)=(0,0)$ and $u+\kappa w = -2$. We here present the nonlinear Dirac Hamiltonian derived from other gapless points.

In the case of $u+\kappa w = 2$, the nonlinear band closes a gap at $(k_x,k_y)=(\pi,\pi)$. Then, the low-energy expansion around these parameters becomes
\begin{eqnarray}
i\partial_t \psi(\mathbf{k},\mathbf{r}) &=& \hat{H}(\psi(\mathbf{k},\mathbf{r})) \psi(\mathbf{k},\mathbf{r}), \\
\hat{H}(\psi(\mathbf{k},\mathbf{r})) &=& \left(
  \begin{array}{cc}
   u-2 + \kappa (|\psi_1(\mathbf{k},\mathbf{r})|^2 + |\psi_2(\mathbf{k},\mathbf{r})|^2) & -k_x - ik_y \\
   -k_x +ik_y & -u+2 - \kappa (|\psi_1(\mathbf{k},\mathbf{r})|^2 + |\psi_2(\mathbf{k},\mathbf{r})|^2)
  \end{array}
  \right). \nonumber\\ \label{nonlinearDirac-wavenumber2}
\end{eqnarray}
Compared to Eq.~(7) in the main text, this nonlinear Dirac Hamiltonian has different signs in the wavenumber-dependent terms and a different value in the mass term $m'=u-2$.

In the case of $u+\kappa w = 2$, the nonlinear QWZ model has two gapless points at $(k_x,k_y)=(0,\pi)$, $(\pi,0)$. Expanding the state-dependent Hamiltonian of the nonlinear QWZ model around $(k_x,k_y)=(0,\pi)$, we obtain
\begin{eqnarray}
i\partial_t \psi(\mathbf{k},\mathbf{r}) &=& \hat{H}(\psi(\mathbf{k},\mathbf{r})) \psi(\mathbf{k},\mathbf{r}), \\
\hat{H}(\psi(\mathbf{k},\mathbf{r})) &=& \left(
  \begin{array}{cc}
   u + \kappa (|\psi_1(\mathbf{k},\mathbf{r})|^2 + |\psi_2(\mathbf{k},\mathbf{r})|^2) & k_x - ik_y \\
   k_x +ik_y & -u - \kappa (|\psi_1(\mathbf{k},\mathbf{r})|^2 + |\psi_2(\mathbf{k},\mathbf{r})|^2)
  \end{array}
  \right). \nonumber\\ \label{nonlinearDirac-wavenumber3}
\end{eqnarray}
When we expand the state-dependent Hamiltonian of the nonlinear QWZ model around $(k_x,k_y)=(\pi,0)$, we obtain
\begin{eqnarray}
i\partial_t \psi(\mathbf{k},\mathbf{r}) &=& \hat{H}(\psi(\mathbf{k},\mathbf{r})) \psi(\mathbf{k},\mathbf{r}), \\
\hat{H}(\psi(\mathbf{k},\mathbf{r})) &=& \left(
  \begin{array}{cc}
   u + \kappa (|\psi_1(\mathbf{k},\mathbf{r})|^2 + |\psi_2(\mathbf{k},\mathbf{r})|^2) & -k_x + ik_y \\
   -k_x -ik_y & -u - \kappa (|\psi_1(\mathbf{k},\mathbf{r})|^2 + |\psi_2(\mathbf{k},\mathbf{r})|^2)
  \end{array}
  \right). \nonumber\\ \label{nonlinearDirac-wavenumber4}
\end{eqnarray}
While these nonlinear Dirac Hamiltonians have different signs in their wavenumber-dependent terms, the nonlinear Chern numbers are determined by the signs of mass terms similar to Eq.~(8) in the main text. The sum of the nonlinear Chern numbers of four nonlinear Dirac Hamiltonians is equal to that of the nonlinear QWZ model.

\begin{figure*}
  \includegraphics[width=140mm,bb=0 0 660 220,clip]{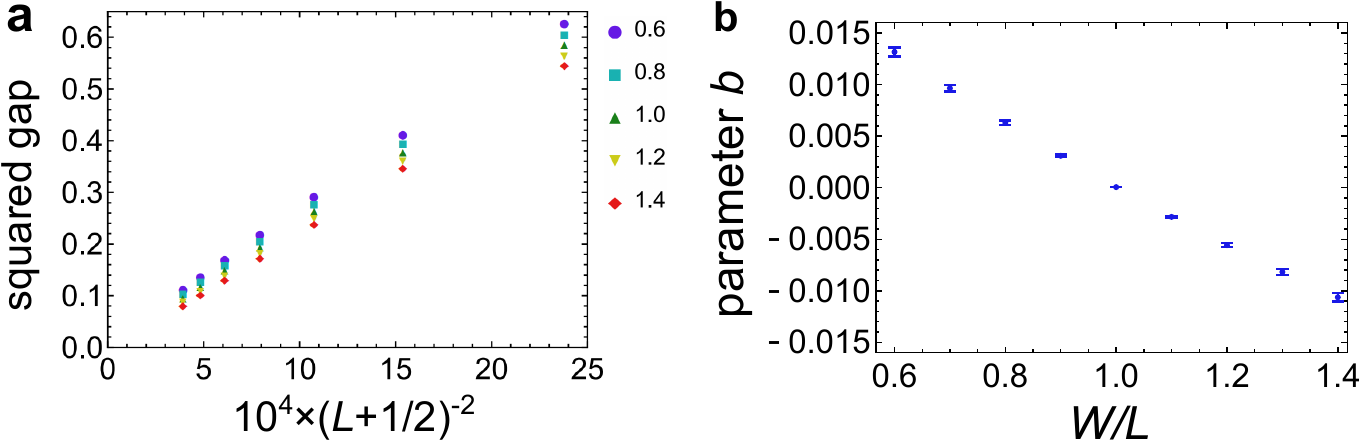}
  \caption{\label{supplefig3} {\bf Recovery of  the bulk-boundary correspondence at the continuum and thermodynamic limit.} {\bf a,} Band gaps at each size and strength of nonlinearity. The squared gaps versus $10^4/(L+1/2)^2$ ($L$ is the size of the system) are plotted. The legend shows the correspondence between the color and the strength of nonlinearity $\kappa w$. The other parameters used are $h=0.01$ and $m=-1$. We can confirm that the gaps decrease as the size become larger. {\bf b,} Squared gaps at the thermodynamic limit estimated from the least-square fitting. We fit the squared gaps with the function $a/(L+1/2)^2+b$ and plot the obtained $b$'s at different amplitudes. $W/L=1.0$ is the transition point, where the band gap is closed and the gapless edge modes appear.}
\end{figure*}

\subsection{System-size dependence of nonlinear band gaps in the nonlinear QWZ Hamiltonian in the continuum limit.}
We analytically estimate the system-size dependence of nonlinear band gaps in the nonlinear QWZ model (Eq.~(4) in the main text) in the continuum limit at the critical amplitude where the nonlinearity-induced topological phase transition occurs. We impose the open boundary condition in the $x$ direction and assume the uniform eigenvector in the $y$ direction. Then, the eigenvalue equation of the discretized system is equivalent to that of the nonlinear Su-Schriefer-Heeger (SSH) model \cite{Su1979,Chen2014,Hadad2016,Tuloup2020,Ezawa2021,Lo2021,Zhou2022},
\begin{eqnarray}
E\vec{\psi}(2x-1) &=& \frac{1}{h}\vec{\psi}(2x-2) + \left[m-\frac{1}{h}+\kappa (|\vec{\psi}(2x-1)|^2+|\vec{\psi}(2x)|^2)\right]\vec{\psi}(2x) \\
E\vec{\psi}(2x) &=& \left[m-\frac{1}{h}+\kappa (|\vec{\psi}(2x-1)|^2+|\vec{\psi}(2x)|^2)\right]\vec{\psi}(2x-1) + \frac{1}{h} \vec{\psi}(2x+1)
\end{eqnarray}
under a properly chosen unitary transformation of the effective Hamiltonian. We use the same parameters $m$, $\kappa$, and $h$ as those in Eq.~(7) in the main text. At the critical amplitude $|\vec{\psi}(2x-1)|^2+|\vec{\psi}(2x)|^2 = |m|/\kappa$, the strengths of intercell and intracell hoppings of the nonlinear SSH model become the same on average, we can estimate the band gap of the discretized nonlinear Dirac Hamiltonian with $L$ unit cells from that of a simple one-dimensional chain with $2L$ lattice points. The eigenvector of the one-dimensional chain is described by sine curves $\psi(x) = \sin(kx)$. To satisfy the open boundary conditions $\psi(0) = 0$ and $\psi(2L+1)=0$, $k$ must be a multiple of $2\pi/(2L+1)$, i.e., $k=2n\pi /(2L+1)$ with $n$ being an integer $n=1,\cdots,2L$. The corresponding eigenvalues are $E = a(2+\cos(2n\pi /(2L+1)))$ with $a$ being a real constant, and thus we estimate the band gap around $E=0$ as $2\pi a /(2L+1) + \mathcal{O}((2L+1)^{-2})$. This estimation indicates that the nonlinear Dirac Hamiltonian is gapless at the critical amplitude in the thermodynamic limit, which is consistent with the bulk dispersion of the nonlinear Dirac Hamiltonian.

We also numerically estimate the nonlinear band gaps in the thermodynamic limit at different amplitudes. We calculate the minimums of the absolute values of eigenvalues at different amplitudes and system sizes. Then, we fit the squares of the minimums by $a/(2L+1)^2+b$, where $a$ and $b$ are the fitting parameters. As we discussed in the previous paragraph, $b$ should become zero at the critical amplitude, while $b$ should be positive below the critical amplitude, which indicates the existence of the band gap in the thermodynamic limit. Supplementary Figure \ref{supplefig3} shows the obtained fitting parameters $b$ at different amplitudes $w$. As we expect, we obtain positive $b$ when the amplitudes are smaller than the critical one, and $b$ is zero at the critical amplitude. Since the fitting function does not completely capture the finite-size scaling of spectrum gaps at noncritical amplitudes, $b$ becomes negative at larger amplitudes than the critical one. However, this numerical result still indicates that the nonlinear QWZ model is gapped (gapless) when the nonlinear Chern number is zero (nonzero).

We note that the nonlinear Dirac Hamiltonian obtained from the low-energy effective theory of the nonlinear QWZ model is different from that in previous studies \cite{Bomantara2017,Smirnova2019} because they have considered the nonlinear terms with the same sign at the first and second components, while we consider those with the opposite signs. Since the difference of the on-site terms corresponds to the mass of the Dirac fermion, the nonlinear on-site terms with the opposite signs can more easily close the nonlinear band gap. Thus, the nonlinear Dirac Hamiltonian considered in this paper exhibit the nonlinearity-induced topological phase transition that has been unexplored in the present studies.
\begin{figure*}
  \includegraphics[width=140mm,bb=0 0 650 250,clip]{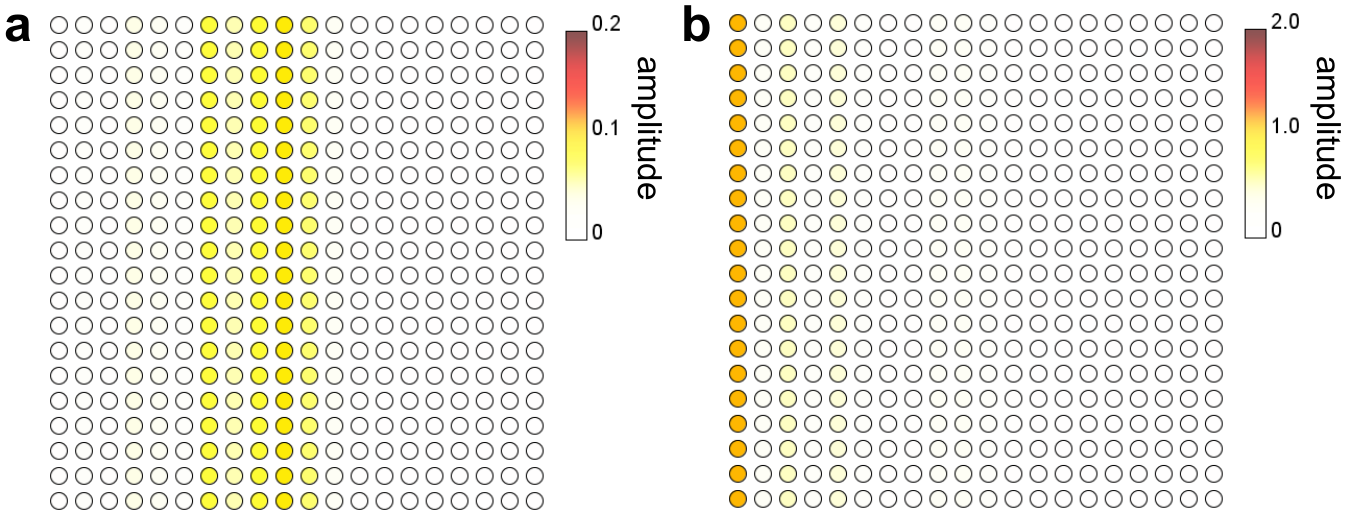}
  \caption{\label{supplefig4} {\bf The existence and absence of localized modes in a stronger nonlinear lattice system.} {\bf a,b,} We calculate the dynamics of the prototypical lattice model of a nonlinear Chern insulator. The color of each site shows the snapshot of the amplitude distribution at the time $t=1$. The parameters used are $u=-2.1$ and $\kappa=1$, and the amplitudes of the initial state differ in two numerical calculations. {\bf a,} We confirm the absence of the localized state at small amplitudes. The amplitude of the initial state is $w=0.03$. {\bf b,} A long-lived localized state is obtained in the stronger nonlinear regime. The amplitude of the initial state is $w=3.0$. The emergence of the localized state indicates the amplitude-dependent topology of nonlinear systems.}
\end{figure*}

\subsection{Nonlinear edge modes in stronger nonlinear lattice systems.}
While in the main text, we analyzed the topology of stronger nonlinear systems in the continuum limit, edge modes can also appear in stronger nonlinear lattice systems. However, there are subtle problems in the bulk-boundary correspondence of such lattice systems as we numerically show below. Specifically, the finite-size effect and the jumps of amplitudes can alter the gapless points at which gapless edge modes can appear. 

\begin{figure*}
  \includegraphics[width=140mm,bb=0 0 790 290,clip]{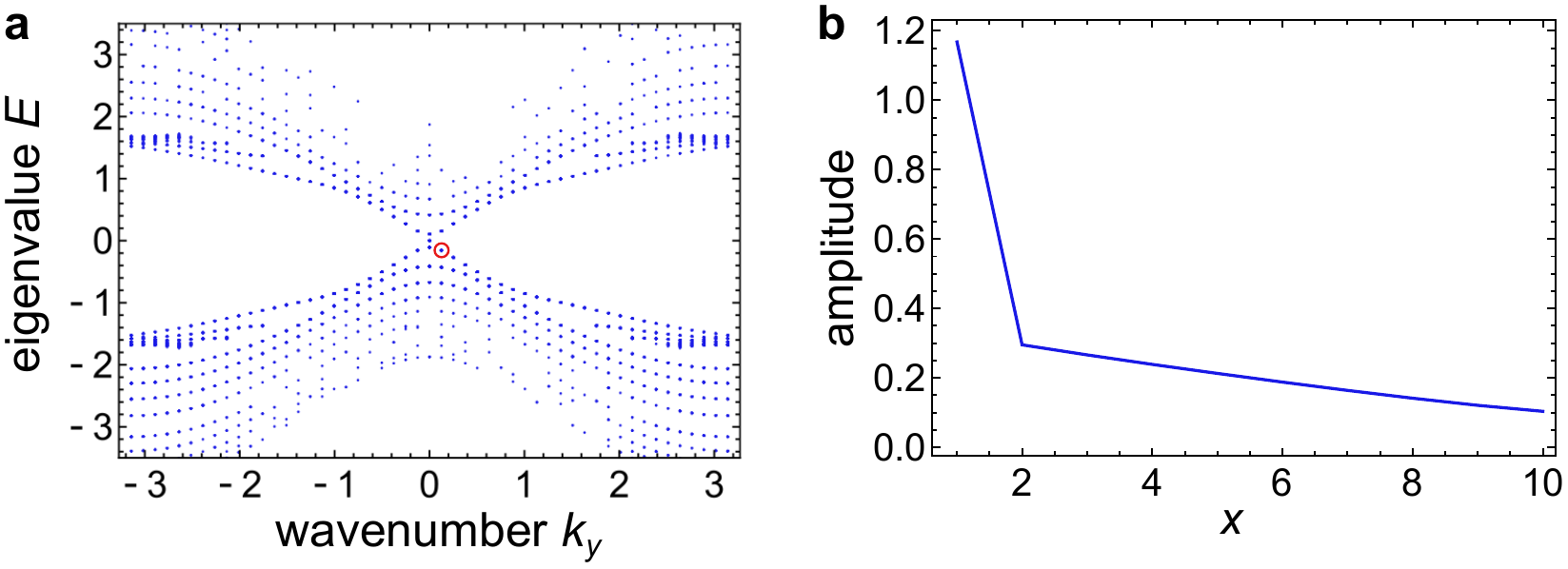}
  \caption{\label{supplefig5} {\bf Nonlinear band structure in a stronger nonlinear regime.} {\bf a,} We calculate the nonlinear band structure of the nonlinear QWZ model (Eq.~(4) in the main text) in a $10\times1$ supercell structure under the open boundary condition in the $x$ direction and the twisted boundary condition in the $y$ direction. We use the parameters $u=-2.1$ and $\kappa w=3$, which correspond to the parameters in Supplementary Fig.~\ref{supplefig4} and exhibit a nonzero nonlinear Chern number $C_{\rm NL} = -1$. We confirm the existence of the gapless modes as predicted from the nonzero nonlinear Chern number. We note that some bulk modes around $k=0$ disappear in this figure due to the limitation of the numerical method {\bf b,} We plot the amplitude distribution of a gapless mode. Its eigenvalue corresponds to that encircled by the red circle in panel {\bf a}. We can confirm the localization at the edge of the sample as in conventional linear topological edge modes.
}
\end{figure*}

To confirm the existence of gapless edge modes in stronger nonlinear lattice systems, we again use the nonlinear QWZ model (Eq.~(4) in the main text). Since the Chern number of the lattice model depends on the amplitude $w$ as discussed in Fig.~2b in the main text, we can expect the emergence of the nonlinearity-induced topological phase transition, where topological edge modes appear at the critical amplitude. In fact, numerically calculating the dynamics of the nonlinear QWZ model above the critical amplitude, we obtain a long-lived localized state shown in Supplementary Fig.~\ref{supplefig4}. Here, we conduct the Runge-Kutta simulation of the lattice model as in Fig.~2 in the main text, setting the time step as $dt=0.005$ and parameters as $u=-2.1$ and $\kappa=1$. 

\begin{figure*}
  \includegraphics[width=140mm,bb=0 0 1245 305,clip]{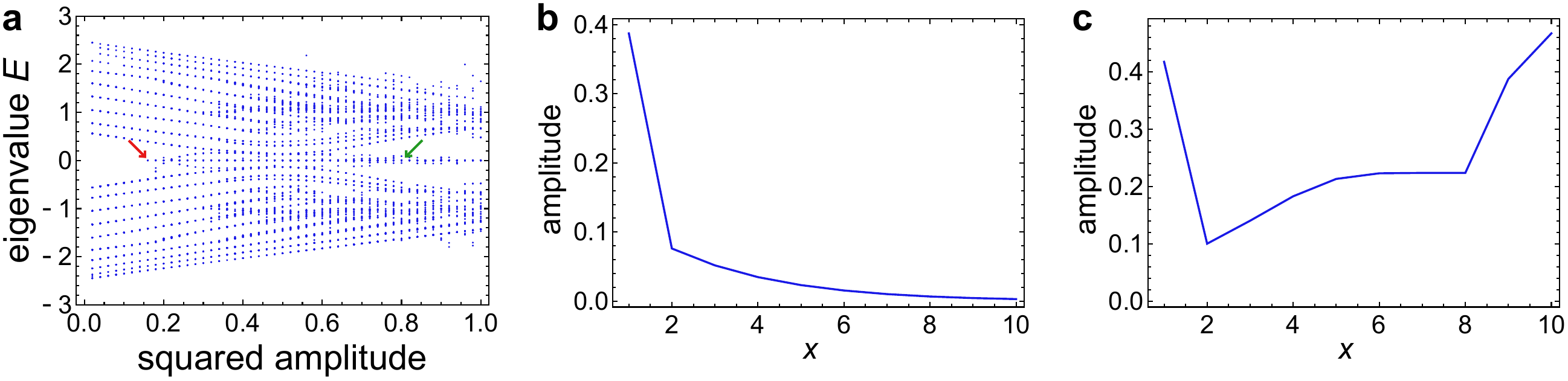}
  \caption{\label{supplefig6} {\bf Nonlinear eigenvalues and the existence of anomalous gapless modes in a stronger nonlinear regime.} {\bf a,} Amplitude dependence of nonlinear eigenvalues at $k_y=0$. We calculate the nonlinear eigenvalues of the nonlinear QWZ model (Eq.~(4) in the main text) in a $10\times1$ supercell structure at $u=-2.5$ and $\kappa=10$. The transition point is $w=0.5$ with $w$ being the squared amplitude, while gapless edge modes appear around $w\simeq 0.15$ as is pointed out by the red arrow. We note that some eigenvalues disappear at large amplitudes due to the limitation of the numerical technique. {\bf b,} Amplitude distribution of an anomalous gapless mode. We confirm the localization of the gapless mode at the edge of the sample. The eigenvalue corresponds to that pointed by the red arrow on the left side of panel {\bf a}. {\bf c,} Amplitude distribution of a topological gapless mode. We confirm the localization of the topological mode at both edges of the system. One can understand this localized edge mode as the superposition of two edge modes localized at the left and right edges, respectively. The eigenvalue corresponds to that pointed by the green arrow on the right side of panel {\bf a}.
}
\end{figure*}

We also numerically calculate the nonlinear band structure corresponding to the lattice model. Since the self-consistent calculation cannot stably obtain the eigenvalues in the strongly nonlinear regimes, here we instead use the quasi-Newton method to solve the nonlinear eigenvalue problems. We impose the open boundary condition in the $x$ direction and the periodic boundary condition in the $y$ direction and calculate the nonlinear eigenvalues at each phase parameter $k_y$ to obtain the nonlinear band structure. Supplementary Figure \ref{supplefig5} shows the nonlinear band structure at the same parameters in Supplementary Fig.~\ref{supplefig4}. One can confirm the existence of the gapless modes. We note that if the amplitude is smaller than that considered in Supplementary Figs.~\ref{supplefig4}, \ref{supplefig5}, there are no gapless edge modes, which indicates that the strongly nonlinear lattice system exhibits the nonlinearity-induced topological phase transition.

The numerical results in Supplementary Figs.~\ref{supplefig4}, \ref{supplefig5} indicate the existence of topological edge modes in stronger nonlinear lattice systems, while one can see the inconsistency between the parameters where edge modes appear and the nonlinear Chern number is changed. We confirm such inconsistency in the numerical calculations of the eigenvalues at $k_y=0$ and different amplitudes. Supplementary Figure \ref{supplefig6}a shows the numerical results at the parameters $u=-2.5$ and $\kappa=10$ and various amplitudes. In this case, the nonlinear Chern number is changed at the amplitude $w=0.5$, while the nonlinear band becomes gapless around $w=0.15$, which implies the breakdown of the bulk-boundary correspondence. We check that the anomalous gapless modes in Supplementary Fig.~\ref{supplefig6}b are also localized at the edge of the sample as in topological gapless modes in Supplementary Fig.~\ref{supplefig6}c. 

The emergence of the anomalous gapless modes is induced by the discontinuous change of the amplitude. At the amplitude $w=0.15$, the local on-site term at $x=0$ can become $u+1+\kappa|\Psi(x=1)|^2=0$, which leads to the existence of perfectly localized modes exhibiting a sudden decrease in their amplitude. Comparing the nonlinear QWZ model and the nonlinear Dirac Hamiltonian (Eq.~(7) in the main text), a sudden change of the amplitude is unphysical in the nonlinear Dirac Hamiltonian. However, the effect of the momentum term becomes much stronger than that of the mass term at the corresponding parameter region, we need to take a small lattice constant to approximate the nonlinear Dirac Hamiltonian by the nonlinear QWZ model. Therefore, using the same lattice constant as in the weakly nonlinear regime, one cannot accurately calculate the high-frequency components of the wavefunction in this parameter region.

\subsection{Exact solutions of the perfectly localized edge modes in the nonlinear QWZ model.}
As discussed in the previous section, there can be perfectly localized modes unique to lattice systems. We here derive the exact solutions of such perfectly localized modes in the nonlinear QWZ model (Eq. (4) in the main text). We consider the localized modes described in the following ansatz,
\begin{equation}
\Psi_1(1,y) = e^{ik_y y}\psi_1,\ \Psi_2(1,y) = e^{ik_y y} \psi_2,
\end{equation}
and
\begin{equation}
\Psi_j(x,y) = 0,
\end{equation}
with $x>1$ and $j=1,2$. We here seek a nonlinear eigenvector whose eigenenergy is $E=k_y$. Then, the nonlinear eigenequation is explicitly given by
\begin{eqnarray}
 \left[ u + 1 + \kappa (|\psi_1|^2+|\psi_2|^2) \right] \psi_1 &=& 0, \\
 \left[ u + 1 + \kappa (|\psi_1|^2+|\psi_2|^2) \right] \psi_2 &=& 0, \\
 \frac{-\psi_1 - i \psi_2}{2} &=& 0, \\
 \frac{\psi_2 - i \psi_1}{2} &=& 0. 
\end{eqnarray}
Solving these equations, we obtain 
\begin{equation}
\left(
  \begin{array}{c}
   \psi_1 \\
   \psi_2
  \end{array}
  \right) = \sqrt{\frac{-u-\cos k_y}{2\kappa}}\left(
  \begin{array}{c}
   1 \\
   i
  \end{array}
  \right),
\end{equation}
under the condition of $(-u-\cos k_y)/\kappa>0$. The squared amplitude of this perfectly-localized mode is
\begin{equation}
w\left( k_{y}\right)L =\left\vert \psi _{1}\right\vert ^{2}+\left\vert \psi_{2}\right\vert ^{2}=\frac{-u-\cos k_{y}}{\kappa }. \label{wky}
\end{equation}
Then, the condition for the emergence of the perfectly localized edge state reads
\begin{equation}
\kappa w = -u-\cos k_y. \label{line_perfectly_localize}
\end{equation}
Such positive $w$ and wavenumber $k_y$ exist when $u$ satisfies $u<1$ in the case of positive $\kappa>0$ and $u>-1$ in the case of negative $\kappa<0$. We note that the light-blue line in Fig.~6d (e) in the main text represents the solution of Supplementary Eq.~\eqref{line_perfectly_localize} at $k_y=0$ ($k_y=\pi$). Therefore, the result of the quench dynamics in Fig.~6 in the main text agrees with these exact edge solutions.

\subsection{Approximated solutions of the trap phase in strongly nonlinear systems.} 
In the strongly nonlinear regime, where the hopping terms are negligible compared with the Kerr nonlinearity, Supplementary Equation \eqref{model-real} is written in the form of
\begin{eqnarray}
i\frac{d\Psi _{1}}{dt} &=&\kappa\left( \left\vert \Psi _{1}\right\vert^{2}+\left\vert \Psi _{2}\right\vert ^{2}\right) \Psi _{1}, \\
i\frac{d\Psi _{2}}{dt} &=&-\kappa\left( \left\vert \Psi _{1}\right\vert^{2}+\left\vert \Psi _{2}\right\vert ^{2}\right) \Psi _{2}.
\end{eqnarray}
By inserting an ansatz 
\begin{equation}
\Psi _{1}=Re^{i\Theta _{1}t},\qquad \Psi _{2}=Re^{i\Theta _{2}t},
\end{equation}
we obtain
\begin{equation}
\frac{d\Theta _{1}}{dt}=-2\kappa R^{2},\qquad \frac{d\Theta _{2}}{dt}=2\kappa R^{2},
\end{equation}
whose solutions read
\begin{equation}
\Psi _{1}=Re^{-2i\kappa R^{2}t},\qquad \Psi _{2}=Re^{2i\kappa R^{2}t}.
\end{equation}
By using the condition
\begin{equation}
\left\vert \Psi _{1}\right\vert ^{2}+\left\vert \Psi _{2}\right\vert ^{2}=w,
\end{equation}
we have
\begin{equation}
R=\sqrt{w/2},
\end{equation}
which leads to the solutions for the trap phase 
\begin{equation}
\Psi _{1}=\sqrt{\frac{w}{2}}e^{-i\kappa wt},\qquad \Psi _{2}=\sqrt{\frac{w}{2}}e^{i\kappa wt},
\end{equation}
where the amplitude is independent of time.

\begin{figure*}
  \includegraphics[width=80mm,bb=0 0 285 215,clip]{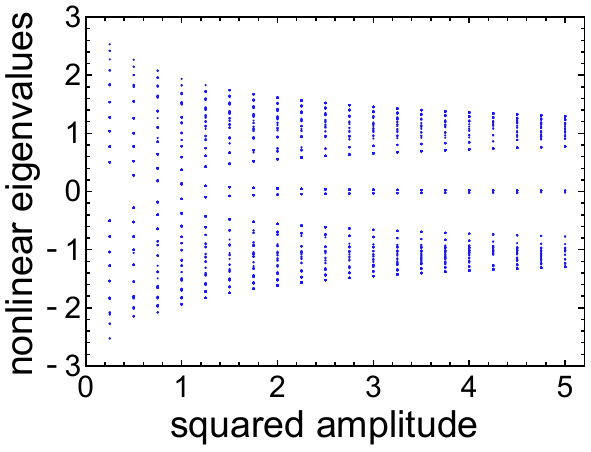}
  \caption{\label{supplefig7} {\bf Recovery of the bulk-boundary correspondence in the infinite-amplitude limit.} We calculate the nonlinear eigenvalues of the modified model of a nonlinear Chern insulator in Supplementary Eq.~\eqref{modified_model}. We impose the open boundary condition in the $x$ direction and the twisted boundary condition in the $y$ direction and plot the nonlinear eigenvalues at $k_y=0$. We use the parameters $u=-3$, $\kappa=2$, and $c=10$. The nonlinear Chern number becomes $C_{\rm NL} = -1$ around the squared amplitude $w\simeq1$ while the band gap is not completely closed at that point. However, one can confirm that the band gap becomes smaller as the amplitude becomes larger, which indicates the existence of genuinely gapless modes and the recovery of the bulk-boundary correspondence in the infinite-amplitude limit.}
\end{figure*}

\subsection{Bulk-boundary correspondence under saturated nonlinearity.}
In strongly nonlinear lattice systems, the bulk-boundary correspondence remains some subtle problems as discussed in Supplementary Note 5, while there are situations that can recover the bulk-boundary correspondence. Specifically, one can confirm the bulk-boundary correspondence under saturated nonlinearity where the strength of the nonlinear term converges in the infinite-amplitude limit. To check this, we construct and analyze the modified nonlinear QWZ model, whose dynamics is described as 
\begin{eqnarray}
&{}& i\frac{d}{dt}\Psi_j(x,y) \nonumber\\
&=& \sum_k \{(\sigma_z)_{jk} [2u\Psi_k(x,y)+\Psi_k(x+1,y)+\Psi_k(x-1,y) +\Psi_k(x,y+1)+\Psi_k(x,y-1)]/2 \nonumber\\
&{}& + (\sigma_x)_{jk}(\Psi_k(x+1,y)-\Psi_k(x-1,y))/2i + (\sigma_y)_{jk}(\Psi_k(x,y+1)-\Psi_k(x,y-1))/2i\} \nonumber\\
&{}& - (-1)^j \kappa \frac{c(|\Psi_1(x,y)|^2+|\Psi_2(x,y)|^2)}{ 1 + c( |\Psi_1(x,y)|^2+|\Psi_2(x,y)|^2 )}\Psi_j(x,y), \label{modified_model}
\end{eqnarray}
This model has different nonlinear terms from those of the nonlinear QWZ model (Eq.~(4) in the main text), and the nonlinear terms converge to $- (-1)^j \kappa$ in the infinite-amplitude limit $w=|\Psi_1(x,y)|^2+|\Psi_2(x,y)|^2 \rightarrow\infty$. Therefore, if we assume $-2<u+\kappa<0$, the nonlinear Chern number becomes $\lim_{w\rightarrow \infty} C_{\rm NL} = -1$ in the infinite-amplitude limit. 

We numerically confirm the emergence of gapless edge modes corresponding to the nonzero nonlinear Chern number in the infinite-amplitude limit. We impose the open boundary condition in the $x$ direction and the twisted boundary condition in the $y$ direction. Then, we focus on the eigenvalues of the phase parameter $k_y=0$, because the nonlinear band structure is symmetric under $k_y\rightarrow -k_y$ and thus can be gapless at symmetric points $k_y=0, \pi$. Supplementary Figure \ref{supplefig7} shows the eigenvalues of Supplementary Eq.~\eqref{modified_model} at $k_y=0$ and different amplitudes. One can confirm that the size of the bandgap converges to zero as the amplitude becomes larger, which indicates the emergence of genuinely gapless modes. Therefore, the modified nonlinear QWZ model exhibits the bulk-boundary correspondence in the infinite-amplitude limit with saturated nonlinearity.

In general, if the nonlinear Chern number converges to a nonzero value in the infinite-amplitude limit, the lattice system exhibits the corresponding number of gapless edge modes. We can intuitively understand this recovery of the bulk-boundary correspondence as follows; $\lim_{w\rightarrow\infty} C_{\rm NL} = c$ means $C_{\rm NL} = c$ for any $w$ larger than a threshold $w_0$. Since almost all components of the nonlinear eigenvector are larger than $w_0/L$ (with $L$ being the system size) in the infinite-amplitude limit, inhomogeneity induced by nonlinear terms in the nonlinear eigenvalue problem does not affect the topological property of the nonlinear system. Therefore, the nonlinear Chern number can predict the existence and absence of gapless edge modes in the infinite-amplitude limit. We note that the nonlinear term in the model analyzed in this section is comparable to the linear term even in the infinite-amplitude limit. Therefore, the nonlinearity is moderate and the existence of the edge modes does not contradict some previous papers \cite{Ezawa2021,Ezawa2022} that have revealed strong nonlinearity induces the bulk-localization and thus eliminates topological edge modes.
\begin{figure*}
  \includegraphics[width=140mm,bb=0 0 790 285,clip]{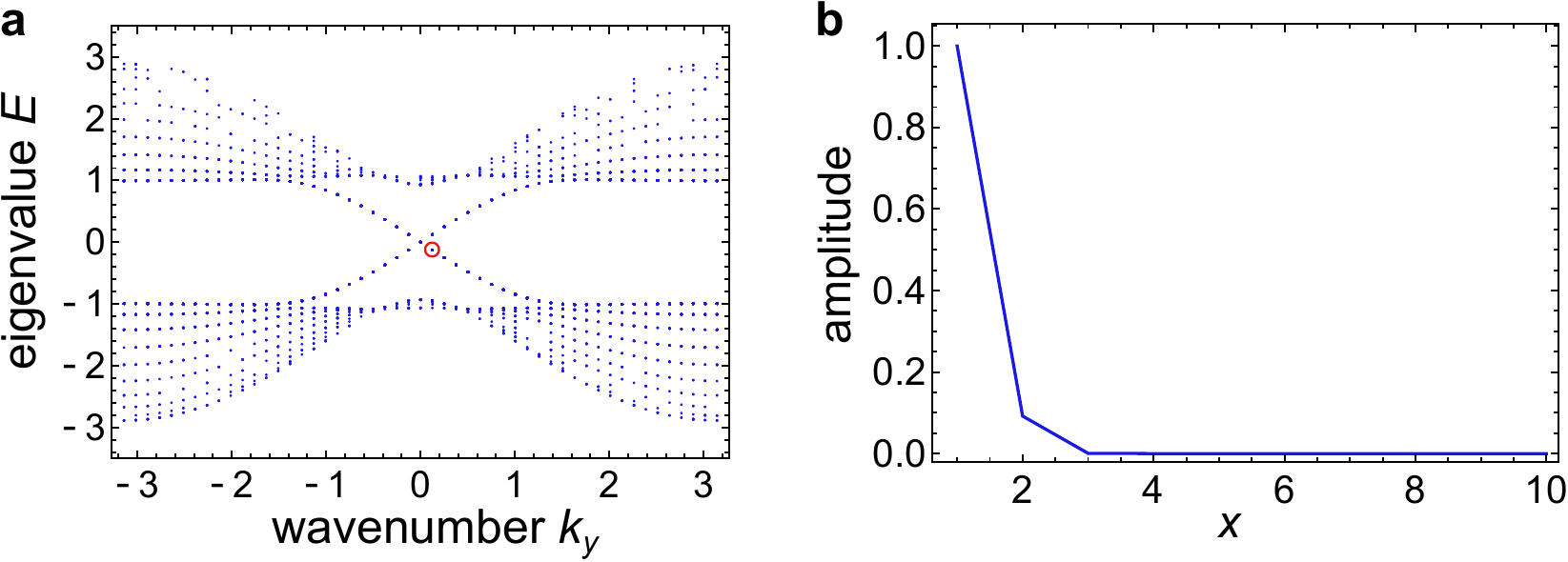}
  \caption{\label{supplefig8} {\bf Nonlinear band structure under the maximum normalization condition.} {\bf a,} We calculate the nonlinear band structure under the weak nonlinearity and the fixed maximum amplitude. We impose the open boundary condition in the $x$ direction and the twisted boundary condition in the $y$ direction. We obtain the gapless edge modes corresponding to the nonzero nonlinear Chern number as in the case of the fixed norm. We use the parameters $u=-1$ and $\kappa w=0.1$. We note that some bulk modes are not obtained due to the limitation of the numerical method. {\bf b,} We show an example of an eigenvector of an edge mode, which corresponds to the red circle in panel {\bf a}. We can check its localization at the edge.}
\end{figure*}
\begin{figure*}
  \includegraphics[width=140mm,bb=0 0 1220 305,clip]{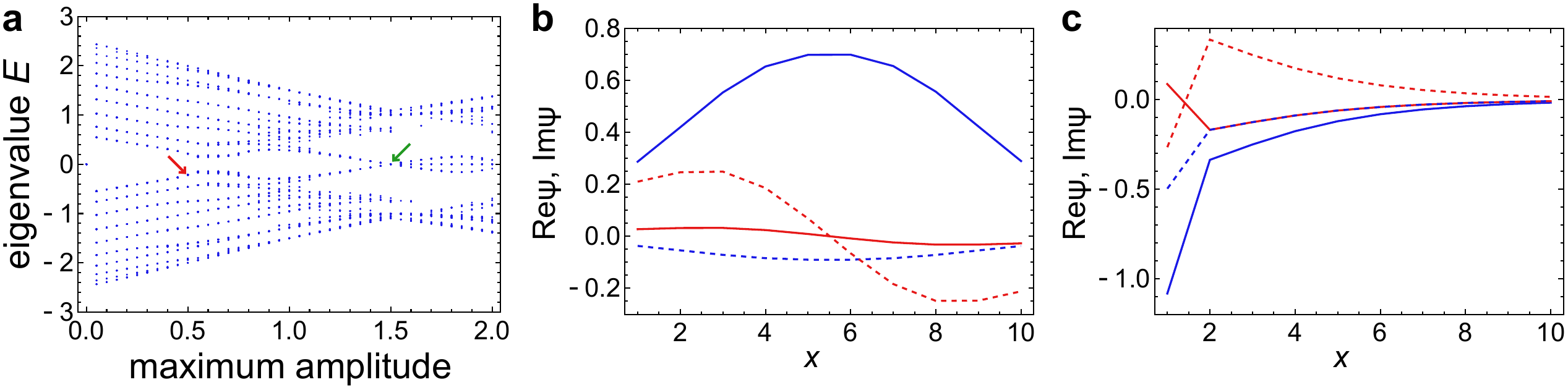}
  \caption{\label{supplefig9} {\bf Nonlinear eigenvalues and eigenvectors under the maximum-based normalization.} {\bf a,} We calculate the nonlinear eigenvalues at various maximum amplitudes. We impose the open boundary condition in the $x$ direction and the twisted boundary condition in the $y$ direction. The transition point of the nonlinear Chern number should correspond to $w=0.5$ with $w$ being the maximum of the squared amplitude, while the nonlinear band has a gap at that parameter. We instead obtain gapless modes at $w=0.5$, which indicates the breakdown of the bulk-boundary correspondence under the maximum-based normalization. {\bf b,} Eigenvector of a bulk mode at $w=0.5$, which is pointed by the red arrow on the left side of panel {\bf a}. The red (blue) curves show the real part of the first (second) component of the wave function, and the red (blue) dashed curves show the imaginary part of the first (second) component. We can confirm that the eigenvector exhibits sine curves, which indicates the absence of edge modes at this amplitude. {\bf c,} Eigenvector of an edge mode at $w=1.5$, which is pointed by the red arrow on the left side of panel {\bf a}. As in panel {\bf b}, the red (blue) curves show the real part of the first (second) component of the wave function, and the red (blue) dashed curves show the imaginary part of the first (second) component. We confirm the localization at the edge of the sample.}
\end{figure*}

\subsection{Effects of definitions of normalization conditions.}
While we fix the $L_2$ norm of the nonlinear eigenvector as the normalization condition, one can consider various types of normalization conditions to restrict the solutions of a nonlinear eigenequation. Specifically, some previous studies \cite{Tuloup2020,Zhou2022} have fixed the maximum absolute value of the components of the nonlinear eigenvector. In this section, we compare the nonlinear band structures under such a normalization condition to those obtained when we fix the $L_2$ norm. Under both normalization conditions, one can confirm the bulk-boundary correspondence in weakly nonlinear systems and in strongly nonlinear continuum systems.

We numerically calculate the eigenvalues of the nonlinear QWZ model (Eq.~(4) in the main text) under the normalization condition of $\max_x (|\psi_1(x)|^2+|\psi_2(x)|^2) = w$. Supplementary Figure \ref{supplefig8} shows the obtained nonlinear band structure under weak nonlinearity. We confirm that the existence and absence of gapless edge modes correspond to the nonzero and zero nonlinear Chern number for each. Under the weak nonlinearity compared to the band gap in the linear limit ($\kappa=0$ in Eq.~(4) in the main text), the nonlinear terms can be considered as the perturbation and thus do not alter the topology of the system independently of the normalization conditions. Therefore, weakly nonlinear systems also exhibit the bulk-boundary correspondence under the normalization condition fixing the maximum of the amplitudes.

We also numerically calculate the nonlinear eigenvalues at $k_y=0$ under the stronger nonlinearity. Supplementary Figure \ref{supplefig9}a shows the obtained spectra in a $10\times 1$ supercell structure under the open boundary condition in the $x$ direction and the twisted boundary condition in the $y$ direction. We use the parameters $u=-2.5$ and $\kappa=1$. Since the mean of the amplitude is less than the maximum (see Supplementary Fig.~\ref{supplefig9}b), the nonlinear band is still gapped at $w=0.5$, where the bulk bands close the gap and the nonlinear Chern number becomes nonzero. Instead, we obtain gapless modes around $w=1.5$ and confirm its localization at the edge of the system as shown in Supplementary Fig.~\ref{supplefig9}c. The emergence of the gapless edge modes indicates the nonlinearity-induced topological phase transition, while the bulk-boundary correspondence is broken under the maximum normalization condition due to similar mechanisms to that under the norm-fixed normalization condition.

\subsection{Exact bulk solutions of the modified nonlinear QWZ model.}
While we focus on the nonlinear QWZ model with nonlinear terms $- (-1)^j \kappa (|\Psi_1(x,y)|^2+|\Psi_2(x,y)|^2) \Psi_i(x,y)$, replacing the nonlinear terms with on-site ones $-(-1)^j|\Psi_j|^2\psi_j$ has no significant effects on the emergence of topological edge modes and the nonlinearity-induced topological phase transitions. We here analytically calculate the bulk solutions of the modified nonlinear QWZ model,
\begin{eqnarray}
i\frac{d}{dt}\Psi_j(x,y) &=& \sum_k \{(\sigma_z)_{jk} [2u\Psi_k(x,y)+\Psi_k(x+1,y)+\Psi_k(x-1,y) +\Psi_k(x,y+1)+\Psi_k(x,y-1)]/2 \nonumber\\
&{}& + (\sigma_x)_{jk}(\Psi_k(x+1,y)-\Psi_k(x-1,y))/2i + (\sigma_y)_{jk}(\Psi_k(x,y+1)-\Psi_k(x,y-1))/2i\} \nonumber\\
&{}& - (-1)^j \kappa |\Psi_j(x,y)|^2 \Psi_j(x,y), \label{model-real-mod}
\end{eqnarray}
which is more feasible in experimental setups using photonics and electrical circuits.

By using the Bloch ansatz, we obtain the following wavenumber-space description,
\begin{eqnarray}
&{}& f(\mathbf{k},\psi(\mathbf{k})) \nonumber\\
&=& \left(
  \begin{array}{cc}
   u + \frac{\kappa w}{2} + c_x+c_y + \frac{\kappa (|\psi_1(\mathbf{k})|^2 - |\psi_2(\mathbf{k})|^2)}{2} & s_x + is_y \\
   s_x - is_y & -\left(u + \frac{\kappa w}{2} +c_x+c_y\right) + \frac{\kappa (|\psi_1(\mathbf{k})|^2 - |\psi_2(\mathbf{k})|^2)}{2}
  \end{array}
  \right) \left(
  \begin{array}{c}
   \psi_1(\mathbf{k}) \\
   \psi_2(\mathbf{k})
  \end{array}
  \right), \nonumber\\ \label{model-wavenumber-mod}
\end{eqnarray}
where $w$ is the amplitude $w=|\psi_1(\mathbf{k}) |^2+|\psi_2(\mathbf{k}) |^2$ and we use the notation $c_i=\cos k_i$, $s_i = \sin k_i$. Since the additional nonlinear term representing the difference of the amplitudes $\kappa(|\psi_1(\mathbf{k})|^2 - |\psi_2(\mathbf{k})|^2)/2$ has the same sign, it only shifts the nonlinear eigenvalues and has no effects on the nonlinear eigenvectors. Therefore, as in the original nonlinear QWZ model, we analytically obtain the nonlinear eigenvalues and eigenvectors as 
\begin{eqnarray}
E_{\pm}(k_x,k_y)&=&\kappa (|\psi_{1\pm}(\mathbf{k})|^2 - |\psi_{2\pm}(\mathbf{k})|^2)/2 + E_{\pm,{\rm NLQWZ}}(k_x,k_y), \label{eigval}\\
E_{\pm,{\rm NLQWZ}}(k_x,k_y) &=& \pm\sqrt{2\cos k_x \cos k_y+2(u+\kappa w/2)(\cos k_x + \cos k_y)+(u+\kappa w/2)^2+2}, \nonumber\\ \\
\left(
  \begin{array}{c}
   \psi_{1\pm}(\mathbf{k}) \\
   \psi_{2\pm}(\mathbf{k})
  \end{array}
  \right) &=& \frac{\sqrt{w}}{c(\mathbf{k})}
  \left(
  \begin{array}{c}
   u+\kappa w/2 + \cos k_x + \cos k_y + E_{\pm,{\rm NLQWZ}}(k_x,k_y) \\
   \sin k_x - i\sin k_y
  \end{array}
  \right).\label{eigvec}
\end{eqnarray}
where $c(\mathbf{k}) = \sqrt{(u+\kappa w/2 + \cos k_x + \cos k_y + E_{\pm,{\rm NLQWZ}}(k_x,k_y))^2+\sin^2 k_x + \sin^2 k_y}$ is a normalization constant. Since the nonlinear eigenvectors are the same as those of the original nonlinear QWZ model at the half of the amplitude (cf.~Eq.~(4) in the main text), we can also exactly calculate the nonlinear Chern number,
\begin{eqnarray}
C_{\rm NL} = \begin{cases}
1 & (0 < u+\kappa w/2 < 2\text{ and gapped}) \\
-1 & (-2 < u+\kappa w/2 < 0\text{ and gapped}) \\
0 & ({\rm otherwise})
\end{cases}.\label{lattice-chern-mod}
\end{eqnarray}
In the calculation of the nonlinear Chern number, we must note that the nonlinear band structure can be gapless in a wide range of amplitude due to the inequivalent shift of the eigenvalues by the additional nonlinear terms $\kappa(|\psi_1(\mathbf{k})|^2 - |\psi_2(\mathbf{k})|^2)/2$.

\subsection{Correspondence between nonlinear eigenvalues and many-body eigenenergies in the mean-field analysis.}
One can obtain nonlinear equations from the mean-field analysis of interacting bosonic systems, which is exemplified by the Gross-Pitaevskii equation of ultracold atoms. In such a mean-field analysis, the nonlinear eigenvector corresponds to the one-particle wavefunction and the nonlinear eigenvector corresponds to the eigenenergy per particle of the interacting bosons. To show this, we describe the derivation of the nonlinear equations of the mean-field theory of interacting bosons \cite{Gross1961,Pitaevskii1961}.

We start from the general many-body Hamiltonian,
\begin{equation}
 H = \sum_i \left[\frac{p_i^2}{2m} + U(x_i)\right] + \sum_{i<j} V(x_i,x_j).
\end{equation}
To derive the mean-field equation, we assume the ansatz,
\begin{equation}
 \Psi(x_1,\cdots,x_N) = \prod_{i=1}^N \psi(x_i).
\end{equation}
Then, the expectation value of the energy is approximated as
\begin{equation}
 \langle \Psi | H | \Psi \rangle = \sum_i \int \psi(x_i)^{\ast} \left[\frac{p_i^2}{2m} + U(x_i)\right] \psi(x_i) dx_i + \sum_{i<j} \int V(x_i,x_j) |\psi(x_i)|^2|\psi(x_j)|^2 dx_i dx_j.
\end{equation}
Finally, the variational method derives the corresponding nonlinear eigenequation,
\begin{equation}
 E_{\rm MB} \psi(x_i) = N\left[\frac{p^2}{2m} + U \right] \psi + \frac{V'N(N-1)}{2} |\psi|^2\psi,
\end{equation}
where we assume the delta-function-like interaction $V\sim V'\delta(x_i-x_j)$ and $E_{\rm MB}$ represents the many-body eigenenergy. One can obtain the corresponding nonlinear eigenequation by dividing the both-hand side by $N$. Therefore, the nonlinear eigenvalue corresponds to the eigenenergy per particle of interacting bosonic systems.

\end{document}